\documentclass[a4paper,fleqn,usenatbib]{mnras}

\usepackage[T1]{fontenc}
\usepackage{ae,aecompl}

\usepackage{amssymb}	
\usepackage[font=small,labelfont=bf]{caption}
\usepackage{amssymb,amsmath}
\usepackage{afterpage} 
\usepackage{amsmath} 
\usepackage{booktabs} 
\usepackage{enumerate} 
\usepackage{rotating} 
\usepackage{graphicx} 
\usepackage{subcaption}
\usepackage{hyperref} 
\usepackage{url}
\usepackage{longtable}
\usepackage{makeidx}
\usepackage{url}
\usepackage{natbib}
\usepackage{multirow}
\usepackage{epstopdf}
\usepackage{caption}
\usepackage[section]{placeins}
\DeclareCaptionFormat{cont}{#1 (cont.)#2#3\par}
\graphicspath{ {figures/} }

\title[NIR and optical counterparts to BHXB]{Quiescent NIR and optical counterparts to candidate black hole X-ray binaries\thanks{Based on observations made with ESO Telescopes at the La Silla Paranal Observatory under programmes 099.A-9025(A) and 1101.D-0105(A).} }

\author[K. M. L\'{o}pez et al.]{
K. M. L\'{o}pez$^{1,2}$\thanks{Contact e-mail: \href{mailto:K.M.Lopez@sron.nl}{K.M.Lopez@sron.nl}},
P. G. Jonker$^{1,2}$,
M. A. P. Torres$^{3,4,1}$,
M. Heida$^{5}$,
A. Rau$^{6}$,
D. Steeghs$^{7}$
\\
\\
$^{1}$SRON Netherlands Institute for Space Research, Sorbonnelaan 2 3584 CA Utrecht, The Netherlands
\\
$^{2}$Department of Astrophysics/IMAPP, Radboud University, P.O. Box 9010, 6500 GL Nijmegen, The Netherlands
\\
$^{3}$Instituto de Astrof\'{i}sica de Canarias, E-38200 La Laguna, Tenerife, Spain
\\
$^{4}$Departamento de Astrof\'{i}sica, Universidad de La Laguna, Astrof\'{i}sico Francisco S\'{a}nchez s/n, E-38206 La Laguna, Tenerife, Spain
\\
$^{5}$Space Radiation Laboratory, California Institute of Technology, Pasadena, CA 91125, USA
\\
$^{6}$Max-Planck-Institut f\"{u}r extraterrestrische Physik, Giessenbachstra{\ss}e 1, 85748 Garching, Germany
\\
$^{7}$Department of Physics, University of Warwick, Coventry CV4 7AL, United Kingdom
}

\date{Accepted XXX. Received YYY; in original form ZZZ}

\pubyear{2018}

\begin{document}
\label{firstpage}
\pagerange{\pageref{firstpage}--\pageref{lastpage}}
\maketitle

\begin{abstract}
We present near-infrared and optical imaging of fifteen candidate black hole X-ray binaries. In addition to quiescent observations for all sources, we also observed two of these sources (IGR J17451-3022 and XTE J1818-245) in outburst. We detect the quiescent counterpart for twelve out of fifteen sources, and for the remaining three we report limiting magnitudes. The magnitudes of the detected counterparts range between $K_s$ = 17.59 and $K_s$ = 22.29 mag. We provide (limits on) the absolute magnitudes and finding charts of all sources. Of these twelve detections in quiescence, seven represent the first quiescent reported values (for MAXI J1543-564, XTE J1726-476, IGR J17451-3022, XTE J1818-245, MAXI J1828-249, MAXI J1836-194, Swift J1910.2-0546) and two detections show fainter counterparts to XTE J1752-223 and XTE J2012+381 than previously reported. We used theoretical arguments and observed trends, for instance between the outburst and quiescent X-ray luminosity and orbital period $P_{orb}$ to derive an expected trend between $\Delta K_s$ and $P_{orb}$ of $\Delta K_s \propto \log P_{orb}^{0.565}$. Comparing this to observations we find a different behaviour. We discuss possible explanations for this result.
\end{abstract}




\section{Introduction}
\label{intro}

An X-ray binary (XRB) is defined as a system in which a neutron star (NS) or a black hole (BH) is accreting matter from its companion star. 
These systems can be classified according to the mass of the donor star, where
we find high-mass X-ray binaries (HMXBs) and low-mass X-ray binaries (LMXBs).
HMXBs are accreting mass through stellar winds or Roche lobe overflow from a massive star (M $\geq$ 10$M_{\odot}$) 
with typical spectral types of O and B for main sequence stars.
The donor companion for LMXBs is typically of K-M spectral type low mass star and has M $\leq$ 1$M_{\odot}$,
which transfers mass by Roche lobe overflow \citep{2003astro.ph..8020C}. XRBs with
spectral type A-F donor stars have also been identified, commonly referred to as intermediate-mass XRBs 
(IMXBs, \citealt{2003ApJ...597.1036P}). These are thought to be the progenitors of some LMXBs \citep{2002ApJ...565.1107P}.

\begin{table*}
\vspace{5mm}
\begin{center}
\caption{Discovery date, position and, when constrained, distance to the BHXBs which are the subject of this paper.}
\label{tab:coord}
\resizebox{\textwidth}{!}{\setlength{\tabcolsep}{12pt}
\begin{tabular}{|lccccccc|}
\toprule
BHXB & Discovery date & R.A. & Dec. & 3-$\sigma$ positional & Galactic & Galactic & Distance\\
&& J2000 & J2000 & uncertainty & latitude $l$ & longitude $b$ & \\
& &(hh:mm:ss) & (dd:mm:ss) & (arcsec) & ($^{\circ}$) & ($^{\circ}$) &(kpc)\\ 
\midrule
Swift J1539.2-6227 & 2008 Nov 24$^a$ & 15:39:12.0 & -62:28:02.3 & 0.5$^p$ & 321.0 & -5.6 & --\\
MAXI J1543-564 & 2011 May 8$^b$ & 15:43:17.3 & -56:24:48.4 & 0.8$^q$ & 325.1 & -1.1 & --\\
XTE J1650-500  & 2001 Sep 5$^c$ & 16:50:01.0 & -49:57:43.6 & 0.6$^r$ & 336.7 & -3.4 & 2.6 $\pm$ 0.7$^{ad}$\\
XTE J1726-476 & 2005 Oct 4$^d$ & 17:26:49.3 & -47:38:24.9 & 1.1$^s$ & 342.2 & -6.9 & --\\
IGR J17451-3022  & 2014 Aug 22$^e$ & 17:45:06.7 & -30:22:43.3 & 0.8$^q$ & 358.7 & -0.6 & --\\
Swift J174510.8-262411 & 2012 Sep 16$^f$ & 17:45:10.8 & -26:24:12.7 & 1.7$^t$ & 2.1 & 1.4 & $< 7^{ae}$\\
XTE J1752-223  & 2009 Oct 23$^g$ &  17:52:15.0951 & -22:20:32.3591 & 0.0008$^u$ & 6.4231 & 2.1143 & 6 $\pm$ 2$^{af}$\\
MAXI J1807+132 & 2017 Mar 13$^h$ & 18:08:07.6 & +13:15:04.6 & 2.3$^v$ & 40.1 & 15.5 & --\\
XTE J1818-245 & 2005 Aug 12$^i$ & 18:18:24.4 & -24:32:18.0 & 1.3$^w$ & 7.4 & -4.2 & --\\
MAXI J1828-249 & 2013 Oct 15$^j$ & 18:28:58.1 & -25:01:45.9 & 4.0$^x$ & 8.1 & -6.5 & --\\
MAXI J1836-194 & 2011 Aug 29$^k$ & 18:35:43.4 & -19:19:10.5 & 0.2$^y$ & 13.9 & -5.3 & 7 $\pm$ 3$^{ag}$\\
XTE J1856+053 & 1996 Sep 17$^l$ & 18:56:42.9 & +05:18:34.3 & 0.3$^z$ & 38.2 & 1.2 & --\\
Swift J1910.2-0546 & 2012 May 31$^m$ & 19:10:22.8 & -05:47:56.4 & 1.3$^{aa}$ & 29.9 & -6.8 & --\\
MAXI J1957+032 & 2015 May 11$^n$ & 19:56:39.1 & +03:26:43.7 & 0.9$^{ab}$ & 43.6 & -12.8 & $\sim$ 6$^{ah}$\\
XTE J2012+381 & 1998 May 24$^o$ & 20:12:37.7 & +38:11:01.2 & 1.2$^{ac}$ & 75.4 & 2.2 & --\\
\bottomrule
\multicolumn{8}{l}{{\bf Notes:} References: $^a$\citet{2008ATel.1855....1K}, $^b$\citet{2011ATel.3330....1N}, $^c$\citet{2001AAS...19915907M}, $^d$\citet{2005ATel..624....1T}, $^e$\citet{2014ATel.6451....1C}, $^f$\citet{2012GCN..13775...1C},}\\
\multicolumn{8}{l}{$^g$\citet{2009ATel.2258....1M}, $^h$\citet{2017ATel.10208...1N}, $^i$\citet{2005ATel..578....1L}, $^j$\citet{2013ATel.5474....1N}, $^k$\citet{2011ATel.3611....1N}, $^l$\citet{1996IAUC.6504....2M},}\\
\multicolumn{8}{l}{$^m$\citet{2012ATel.4140....1U}, $^n$\citet{2015ATel.7504....1N,2015ATel.7506....1C}, $^o$\citet{1998IAUC.6920....1R}, $^p$\citet{2011ApJ...735..104K}, $^q$\citet{2014ATel.6533....1C},}\\
\multicolumn{8}{l}{$^r$\citet{2004ApJ...601..439T}, $^s$\citet{2005ATel..628....1M}, $^t$\citet{2012ATel.4380....1R}, $^u$\citet{2011MNRAS.415..306M}, $^v$\citet{2017ATel.10215...1K}, $^w$\citet{2005ATel..589....1R},}\\
\multicolumn{8}{l}{$^x$\citet{2013ATel.5479....1K}, $^y$\citet{2015MNRAS.450.1745R}, $^z$\citet{2007ATel.1072....1T}, $^{aa}$\citet{2012ATel.4144....1R}, $^{ab}$\citet{2015ATel.7524....1R}, $^{ac}$\citet{1998IAUC.6932....2H}, $^{ad}$\citet{2006MNRAS.366..235H}.}\\
\multicolumn{8}{l}{$^{ae}$\citet{2013MNRAS.432.1133M}, $^{af}$\citet{2012MNRAS.423.2656R}, $^{ag}$\citet{2014MNRAS.439.1381R}, and $^{ah}$\citet{2017MNRAS.468..564M}. }\\
\multicolumn{8}{l}{*The coordinates for the counterpart that we identified are R.A. = 17:45:06.654, Dec. = -30:22:43.67.}\\
\end{tabular}}
\end{center}
\end{table*}

Whether an XRB is a (semi) persistent source or shows outburst -- quiescence cycles depends on the
mass transfer rate and the orbital period $P_{orb}$ \citep{1996ApJ...464L.127K}.
Persistent sources typically have X-ray luminosities from $1-100\%$ of the Eddington limit, causing the accretion
disc to dominate the optical spectrum, hiding the companion star in most cases, except a few giant star
mass donors (e.g. \citealt{2005MNRAS.356..621J,2013ApJ...768..185S}). 
On the other hand, X-Ray Transients (XRTs) are characterised by 
episodic outbursts caused by mass transfer instabilities in the accretion disk \citep{2001A&A...373..251D}; between 
these outbursts, they decay back to the quiescent state ($L_X <$ 10$^{33}$ erg s$^{-1}$, e.g. 
\citealt{1998ASPC..137..506G,2000A&A...360..575L,2009MNRAS.392..665G,2008AIPC..983..519J,2013ApJ...773...59P,2016ApJ...826..149B,2017MNRAS.468..564M}), where the optical detection of the donor star is often, 
though not always (e.g. \citealt{2015MNRAS.450.4292T}), possible. This gives the
opportunity to perform radial velocity measurements which allows us to study the 
orbital period evolution and to measure
dynamically the mass of the compact accretor \citep{2014SSRv..183..223C}. 
The latter is the best way to determine its nature since BH and NS systems display
similar outburst properties such as hysteresis patterns in their hardness-intensity diagrams
\citep{2014MNRAS.443.3270M}, though the X-ray timing properties do show differences between NS
and BH accretors (see for instance the overview of \citealt{2010csxs.book.....L}). It is also possible to use the radio
properties as BHs are more radio-loud than NSs \citep{2001MNRAS.324..923F,2003MNRAS.343L..99F}.

XRTs are often first detected at X-ray wavelengths in outburst thanks to all-sky X-ray
monitoring programs, for instance through the RXTE satellite before 2012 \citep{1993A&AS...97..355B}, 
INTEGRAL \citep{2007A&A...466..595K}, Neil Gehrels Swift Observatory \citep{2013ApJS..209...14K} and MAXI 
\citep{2009PASJ...61..999M} missions. Until today, we have 19 dynamically confirmed 
galactic stellar-mass black holes
(\citealt{2014SSRv..183..223C} and references therein, \citealt{2015MNRAS.454.2199M}, \citealt{2016ApJ...825...10T}), 
where 18 of these are in L/IMXBs and one in a HMXB 
(Cyg X-1, \citealt{2011ApJ...742...84O}). 
There is a concentration
towards the Galactic bulge and plane in the spatial distribution of these sources 
(i.e. $340^{\circ} < l < 20^{\circ}$ and $\mid b\mid < 10^{\circ}$, e.g. \citealt{2004MNRAS.354..355J}).
Around 50$\%$ of these 19 sources are located
within about 4.5 kpc from the Sun, which indicates that the dynamical mass measurements are challenging
observationally due to the faintness of the mass donor stars, in part due to the sometimes high interstellar 
extinction towards sources located in the plane/bulge regions.
Additionally, about 40 black 
hole X-ray binary (BHXB) candidates are known, which even though they do not have an estimate
for their mass, present similar outburst properties to already classified BHXBs
(see \citealt{2011BASI...39..409B} for a review). 
The number of BHXB candidates grows at an
approximate rate of 2 objects/year \citep{2016A&A...587A..61C}. 

As mentioned before, the quiescent state of XRTs is the ideal starting point to perform
dynamical studies of these binary systems. The more dynamically confirmed black holes are known, the
more information we have about the mass distribution of BHs in our Galaxy. This is important for
understanding the physics of supernova explosions, the equation of state of nuclear matter \citep{2017arXiv170107450C}
and the survival of interacting binaries, including those that might eventually merge and
produce bursts of graviational wave radiation (e.g. \citealt{2016PhRvL.116f1102A,2016PhRvL.116x1103A}). 
The mass distribution of BHs is
expected to be smooth \citep{2001ApJ...554..548F}; however, observations have shown a gap between 
NSs and BHs in the range 2--5$M_{\odot}$ \citep{2010ApJ...725.1918O,2011ApJ...741..103F}. The
existence of the gap is still under debate. Some argue that the observed distribution may be biased
by selection efects (i.e. \citealt{2010ApJ...725.1918O}) and biases in the mass measurement procedure.
For instance, \citet{2012ApJ...757...36K} claim that there might be a systematic trend in the
inclination angle determinations that lead to an underestimate of the 
inclination and thus, an overestimate of the BH mass (see also \citealt{2017MNRAS.472.1907V}).
Others state that the gap is real and could shed light 
on supernovae explosion models (i.e. \citealt{2012ApJ...757...91B,2012ApJ...749...91F}). 
With Gaia \citep{2016A&A...595A...1G} we will be able to obtain accurate distances and proper motions for several 
XRBs, necessary to determine BH natal kicks (e.g. \citealt{1995ApJ...447L..33V,2004MNRAS.354..355J,2009ApJ...706L.230M,2011ApJ...742...83R,2014ApJ...796....2R,2015MNRAS.453.3341R,2016MNRAS.456..578M}) 
and hence, constrain the formation and evolution of BHXBs.

\begin{table*}
\vspace{5mm}
\begin{center}
\caption{Observing log of the black hole X-ray binary candidates discussed in this paper. All of the sources are observed in quiescence. An * marks sources also observed in outburst.}
\label{tab:sources}
\resizebox{\textwidth}{!}{\begin{tabular}{|lcccccccc|}
\toprule
BHXB & Instrument & Date & On source & 1-$\sigma$ WCS & Photometric & Limiting &Filter & Average\\
candidate & & observed & time & uncertainty$^a$ & zero point$^b$ & magnitude$^c$ & & seeing\\
 & & & (sec) &(mas) & (mag) & & &(\arcsec)\\
\midrule
Swift J1539.2-6227 & GROND & 2017 September 15 & 372 & 130 $\pm$ 20 & 25.61 $\pm$ 0.06& $< 19.6$&$g'$ & 2.2 \\
 &  &  & 372 & 60 $\pm$ 20 & 26.34 $\pm$ 0.03 & $< 20.1$&$r'$ & 1.9 \\
 &  &  & 372 & 50 $\pm$ 15 & 25.41 $\pm$ 0.06 & $< 19.8$&$i'$ & 1.8 \\
&  &  & 372 & 80 $\pm$ 15 & 25.18 $\pm$ 0.03 & $<18.9 $&$z'$ & 1.8 \\
 &  &  & 480 & 125 $\pm$ 15 & 24.32 $\pm$ 0.05 & $< 18.9$&$J$ & 1.8 \\
 &  &  & 480 & 340 $\pm$ 15 & 23.85 $\pm$ 0.07 & $< 16.8$&$H$ & 1.9 \\
 &  &  & 480 & 50 $\pm$ 15 & 24.57 $\pm$ 0.05 & $< 17.6$&$K_s$ & 1.8 \\
MAXI J1543-564 & HAWK-I & 2018 April 10 & 480 & 170 $\pm$ 15 & 27.56 $\pm$ 0.17 & $< 23.2$&$H$ & 0.5 \\
 &  &  & 420 & 270 $\pm$ 15 &  27.31 $\pm$ 0.16 & $< 23.2$&$K_s$ & 0.5 \\
XTE J1650-500 & GROND & 2017 September 14 & 372 & 100 $\pm$ 20 & 25.61 $\pm$ 0.06 & $< 20.3$&$g'$ & 2.0 \\
 &  & & 372 & 100 $\pm$ 20 & 26.34 $\pm$ 0.03 & $< 20.5$&$r'$ & 2.0 \\
&  &  & 372 & 60 $\pm$ 15 & 25.41 $\pm$ 0.06 & $< 19.4$&$i'$ & 1.9 \\
 &  &  & 372 & 120 $\pm$ 15 & 25.18 $\pm$ 0.03 & $< 18.5$&$z'$ & 1.8 \\
 &  &  & 480 & 50 $\pm$ 15 & 24.32 $\pm$ 0.05 & $< 18.4$&$J$ & 1.9 \\
&  &  & 480 & 60 $\pm$ 15 & 23.85 $\pm$ 0.07 & $< 16.8$&$H$ & 2.0 \\
&  &  & 480 & 40 $\pm$ 15 & 24.57 $\pm$ 0.05 & $< 17.8$&$K_s$ & 1.9 \\
XTE J1726-476 & PANIC & 2006 August 3 & 900 & 100 $\pm$ 15 & 25.03 $\pm$ 0.24 &$< $ 20.7 &$J$ & 0.7 \\
 & & & 375 & 200 $\pm$ 15 & 25.38 $\pm$ 0.26 & $< 17.9 $ &$K_s$ & 0.6 \\
IGR J17451-3022 & LIRIS & 2015 April 9$^*$ & 360 & 110 $\pm$ 15 & 24.29 $\pm$ 0.09 & $< 17.5$ &$K_s$ & 0.7 \\
&  & 2016 March 28 & 450 & 80 $\pm$ 15 & 24.41 $\pm$ 0.08 & $< 17.7$&$K_s$ & 0.8 \\
Swift J174510.8-262411 & LIRIS & 2017 April 13 & 1600 & 100 $\pm$ 15 & 24.80 $\pm$ 0.05 & $< 19.0$&$K_s$ & 0.9 \\
XTE J1752-223 & LIRIS & 2017 April 13 & 1600 & 40 $\pm$ 15 & 24.04 $\pm$ 0.07 & $< 19.6$&$K_s$ & 1.0 \\
MAXI J1807+132 & ACAM & 2017 July 11 & 1800 & 120 $\pm$ 20 & 25.88 $\pm$ 0.05 & $< 24.4$&$i'$ & 1.3 \\
XTE J1818-245 & PANIC & 2005 September 12$^*$ & 450 & 80 $\pm$ 15 & 25.39 $\pm$ 0.01 & $< $ 19.3 &$K_s$ & 0.5 \\
 & & 2006 May 8 & 375 & 110 $\pm$ 15 & 25.62 $\pm$ 0.01 & $< $ 17.6 &$K_s$ & 0.6 \\
MAXI J1828-249 & MOSFIRE & 2017 June 15 & 419 & 160 $\pm$ 15 & 27.41 $\pm$ 0.08 & $< 21.9$&$K_s$ & 1.1 \\
MAXI J1836-194 & MOSFIRE & 2017 June 15 & 436 & 60 $\pm$ 15 & 27.51 $\pm$ 0.29 & $< 22.2$&$K_s$ & 1.1 \\
XTE J1856+053 & MOSFIRE & 2017 June 15 & 436 & 50 $\pm$ 15 & 28.69 $\pm$ 0.25 & $< 20.7$&$K_s$ & 1.1 \\
Swift J1910.2-0546 & ACAM & 2015 July 19 & 600 & 110 $\pm$ 15 & 26.25 $\pm$ 0.01 & $< 22.4$& $i'$ &  1.1\\
& & & 600 & 150 $\pm$ 15 & 26.24 $\pm$ 0.01 & $< 22.5$& $i'$ & 1.2\\
& & & 600 & 100 $\pm$ 15 & 26.39 $\pm$ 0.12 & $< 23.6$& $r'$ & 1.3\\
 & LIRIS & 2017 April 13 & 800 & 20 $\pm$ 15 & 24.68 $\pm$ 0.10 & $< 22.1$&$K_s$ & 0.8 \\
MAXI J1957+032 & MOSFIRE & 2017 June 15 & 291 & 80 $\pm$ 15 & 27.63 $\pm$ 0.13 & $< 23.3$&$K_s$ & 0.9 \\
XTE J2012+381 & LIRIS & 2017 April 13 & 2000 & 80 $\pm$ 15 & 24.67 $\pm$ 0.22 & $< 21.5$&$K_s$ & 0.8 \\
\bottomrule
\multicolumn{9}{l}{{\bf Notes:} $^*$Observation in outburst. $^a$Uncertainty with respect to the reference catalog. The first value is a statistical uncertainty, given by the}\\
\multicolumn{9}{l}{{\scshape starlink gaia} tool, while the second value is a systematic uncertainty, and corresponds to the astrometric accuracy of the reference catalog}\\
\multicolumn{9}{l}{(20 mas for UCAC4 and 15 mas for 2MASS). $^b$Photometric zero point derived from the data with respect to the 2MASS catalogue, or VVV}\\
\multicolumn{9}{l}{when possible, (for all the NIR sources), and the Pan-STARRS 1 catalogue (for the optical sources). $^c$Limiting magnitude close to the}\\
\multicolumn{9}{l}{position of the BHXB.}\\
\end{tabular}}
\end{center}
\end{table*}

In this manuscript we report the quiescent magnitudes, or upper limits, of the counterparts to 15 BHXB 
candidates in the near-infrared (NIR) and optical bands. 
For seven of them our
measurements are the first reported to date. We provide also outburst magnitudes for 
two of these 15 sources (XTE J1818-245 and IGR J17451-3022); moreover, we detected for the first time the counterpart for IGR J17451-3022. 
We describe our sample in Section~
\ref{Sample}, the observations in Section~\ref{obs} 
and the data reduction and analysis in
Section~\ref{datared}. Our results are presented in 
Section~\ref{results} and discussed in Section~\ref{discussion},
and we end with the conclusions of our work in 
Section~\ref{conclusions}. We report values taken from literature as found in the original work, i.e.
if uncertainties were originally not given, we report the value without uncertainties
(this includes points without error bars in our plots).

\section{Sample}
\label{Sample}

Our sample consists of 15 BHXB candidates (see Table~\ref{tab:coord} for
discovery dates, coordinates and distances). With the exception of MAXI J1957+032 and MAXI J1807+132,
the sources can be found in either the BlackCAT \citep{2016A&A...587A..61C} or WATCHDOG \citep{2016ApJS..222...15T} catalogues for BHXB candidates. 
Both MAXI J1957+032 and MAXI J1807+132 are neutron star candidates \citep{2017ApJ...851..114R,2017ApJ...850..155S},
reported after we had observed them; however, their spectral features are also consistent with them being BH XRTs
 as no Type I X-ray bursts or pulsation have been found \citep{2017MNRAS.468..564M,2017ATel.10221...1M,2017ATel.10222...1S}.
Thirteen sources are located at latitudes
$\mid b\mid < 7^{\circ}$, whereas MAXI J1957+032 is at $ b \sim$ 13$^{\circ}$ and 
MAXI J1807+132 is at $ b \sim$ 15$^{\circ}$. 
All fifteen sources are observed during quiescence: seven sources in the $K_s$--band
(XTE J2012+381, XTE J1752-223, Swift J174510.8-262411, MAXI J1828-249, MAXI J1836-194, 
MAXI J1957+032, and XTE J1856+053), XTE J1726-476 in the $J$ and $K_s$ bands, MAXI J1543-564 in the $H$ and $K_s$ bands, MAXI J1807+132 in the
$i'$--band, Swift 1910.2-0546 in the $K_s, r', i'$ bands and two sources (XTE J1650-500 and Swift J1539.2-6227) in the $g', r', i', z', J, H, K_s$ bands. 
XTE J1818-245 and IGR J17451-3022
are observed both in quiescence and in outburst, both in the $K_s$--band (see Table~\ref{tab:sources}).

\section{Observations}
\label{obs}

We obtained NIR and optical observations for this project using five different telescopes.
To acquire $JHK_s$--band images we used the William Herschel Telescope with the
Long-slit Intermediate Resolution Infrared Spectrograph (WHT/LIRIS), 
the Keck I telescope on Mauna Kea with the Multi-Object Spectrometer for 
Infra-Red Exploration (Keck/MOSFIRE, \citealt{2010SPIE.7735E..1EM,2012SPIE.8446E..0JM}), the 
Walter Baade Magellan Telescope with
the Persson's Auxiliary Nasmyth Infrared Camera (Magellan/PANIC, \citealt{2004SPIE.5492.1653M}) and
the Very Large Telescope Unit 4 with the High Acuity Wide field K-band Imager (VLT/HAWK-I, \citealt{2004SPIE.5492.1763P}).
To obtain the optical ($r'i'$--band) data we used the William Herschel Telescope with the auxiliary-port camera 
(WHT/ACAM). Finally, we got NIR and optical observations in seven filters ($g', r', i', z', J, H, K_s$) using the 
Gamma-Ray Burst Optical/Near-Infrared Detector (MPI/GROND, 
\citealt{2008PASP..120..405G}) at the
MPI/ESO 2.2 m telescope at the ESO La Silla Observatory.

LIRIS has a pixel scale of 0.25\arcsec\ /pixel and a field of view of 
4.27\arcmin\ $\times$ 4.27\arcmin\ ; MOSFIRE provides a pixel scale of 
0.18\arcsec\ /pixel and a field of view of 6.1\arcmin\ $\times$ 6.1\arcmin\ ; and
PANIC has a pixel scale of 0.125\arcsec\ /pixel and a field of view of 
2.0\arcmin\ $\times$ 2.0\arcmin\ . HAWK-I provides a pixel scale of 0.106\arcsec\ /pixel and a field of view of 7.5\arcmin\ $\times$ 7.7\arcmin\ , ACAM has a pixel scale of 
0.25\arcsec\ /pixel and a circular field of view with a diameter of 8.0\arcmin\ , and GROND provides a pixel scale in the 
optical(NIR) channels of 0.158\arcsec\ (0.60\arcsec\ )/pixel, and a field of view 
in the optical(NIR) channel of of 5.4\arcmin\ $\times$ 5.4\arcmin\ (10\arcmin\ $\times$ 10\arcmin\ ).
The NIR observations with LIRIS, PANIC and MOSFIRE were
performed using multiple repetitions of a 5-position-dither pattern where several images 
were taken at each dither position, while the HAWK-I observations\footnote{ESO program 1101.D-0105(A), PI P.G.Jonker} were done with 8 pointings of 60 s each for the $H$--band and 7 pointings of 60 s each for the $K_s$--band. The optical observation with WHT/ACAM was done 
with 6 different pointings of 300 s each; additionally, the GROND observations\footnote{ESO program 099.A-9025(A), PI A.Rau} in the $g', r', i', z'$ filters were done with 8 pointings of 62 s each at different offset positions, while the observations in the $J, H, K_s$ filters were done with 48 pointings of 10 s, dithered only for the $K_s$--band and at different positions for the $J, H$--bands.

We targeted two sources in the $K_s$--band with PANIC: XTE J1818-245 was observed in 
outburst in September 2005 and then in quiescence in May 2006, and XTE J1726-476 
was observed in quiescence in August 2006 in both $J$ and $K_s$ bands. Five sources were observed
in the $K_s$--band with LIRIS: IGR J17451-3022 was observed in outburst in April 2015
and then in quiescence in March 2016, and four other sources (XTE J2012+381, XTE J1752-223, 
Swift J174510.8-262411 and Swift J1910.2-0546) were observed 
in quiescence in April 2017. The last source, Swift J1910.2-0545, was
also observed in quiescence with ACAM in both $r'$ and $i'$ bands, in July 2015.
We took $K_s$--band images with Keck/MOSFIRE for 
four sources (MAXI J1828-249, MAXI J1836-194, MAXI J1957+032 and XTE J1856+053) 
in quiescence in June 2017, MAXI J1807+132 was observed in quiescence with ACAM 
in the $i'$--band in July 2017, two sources in quiescence (XTE J1650-500 and 
Swift J1539.2-6227) were observed with GROND
in September 2017, and finally, MAXI J1453-564 was observed in quiescence in both $H$ and $K_s$ bands with HAWK-I in April 2018. 
For a full observing log see Table~\ref{tab:sources}, where also the average seeing is provided, as a
measure of the image quality during the observations.

\section{Data reduction and analysis}
\label{datared}

\subsection{Data reduction}
\label{datar}

For the images taken with GROND, the data reduction
was done with the standard tools and methods described in \citet{2008ApJ...685..376K}. 
The PANIC data were processed with {\scshape IRAF} 
scripts implemented by \citet{2004SPIE.5492.1653M} for the reduction of data from
that instrument. These procedures included dark and flat-field correction as well as sky subtraction. 
In the data reduction process, the raw frames were first dark subtracted. 
Master flat-fields were built by combining twilight flat-field frames scaled by the
mode and these were applied to the target images. A sky image was made by masking 
out stars from each set of dithered target frames and was subtracted from the 
associated set of frames. Finally, a single target image was obtained by average 
combining the sky-subtracted images.

The LIRIS, ACAM, MOSFIRE and HAWK-I data were reduced using the data 
reduction software {\scshape theli} \citep{2013ApJS..209...21S}.
In order to flat-field correct the NIR data, we produced a master flat with {\scshape theli} by
median combining the sky flat images taken during twilight.
Additionally, a sky background model is generated by median combining the
observations without correcting for the offsets introduced by the
dithering, which then is subtracted from the individual
data frames. After this, {\scshape theli} detects sources in the images using {\scshape SExtractor} \citep{1996A&AS..117..393B} and obtains astrometric solutions with {\scshape scamp} 
\citep{2006ASPC..351..112B}.
The astrometric solution is obtained by matching the detected
positions to sources from different catalogues.
The global astrometric solution is then used for the coaddition of all the
frames using {\scshape swarp} \citep{2002ASPC..281..228B}.

\subsection{Astrometry}
\label{astrometry}

To improve the accuracy of the global astrometric solution 
of the coadded images, we used the {\scshape starlink}
tool {\scshape gaia}, fitting at least 5 star positions from the 2 Micron All Sky Survey
(2MASS; \citealt{2006AJ....131.1163S}) or from the fourth US 
Naval Observatory CCD Astrograph Catalog (UCAC4, \citealt{2013AJ....145...44Z}), in order
to build a local astrometric solution around the position of the sources.
The rms errors of these fits are indicated in Table~\ref{tab:sources} 
as WCS (World Coordinate System) uncertainties,
where the intrinsic systematic error of the catalog with respect 
to the International Celestial Reference System (ICRS) is also listed 
(15 mas for 2MASS and 20 mas for UCAC4).

\begin{figure*}
     \begin{minipage}{0.4\textwidth}
        \includegraphics[width=\textwidth]{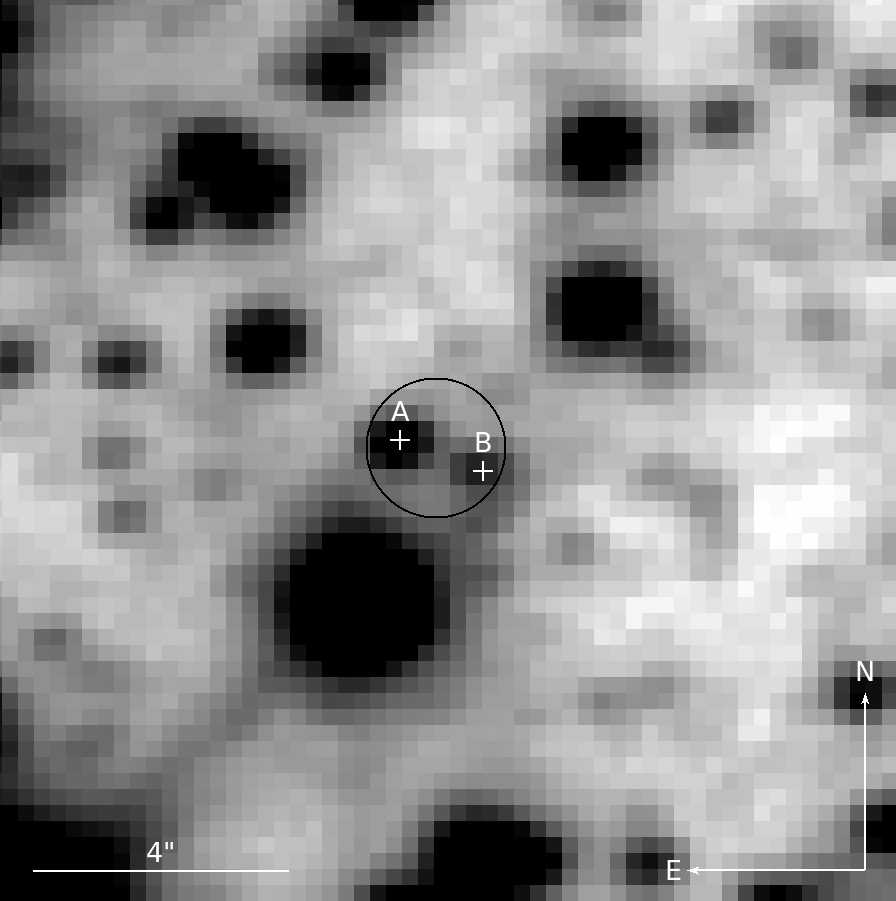}%
        \subcaption{IGR J17451 in outburst}
        \label{fig:igrj17451o}
         \vspace{0.5cm}
    \end{minipage}%
    \begin{minipage}{0.4\textwidth}
        \includegraphics[width=\textwidth]{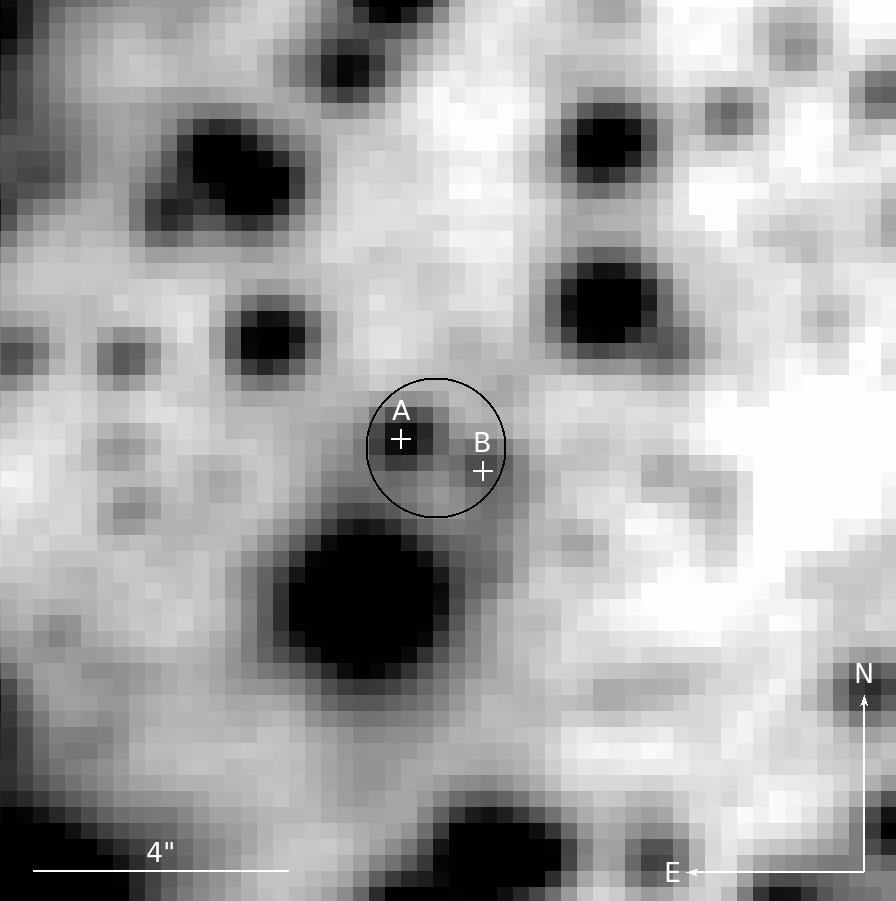}%
       \subcaption{IGR J17451 in quiescence} 
       \label{fig:igrj17451q}
        \vspace{0.5cm}       
    \end{minipage}
    \begin{minipage}{0.4\textwidth}
        \includegraphics[width=\textwidth]{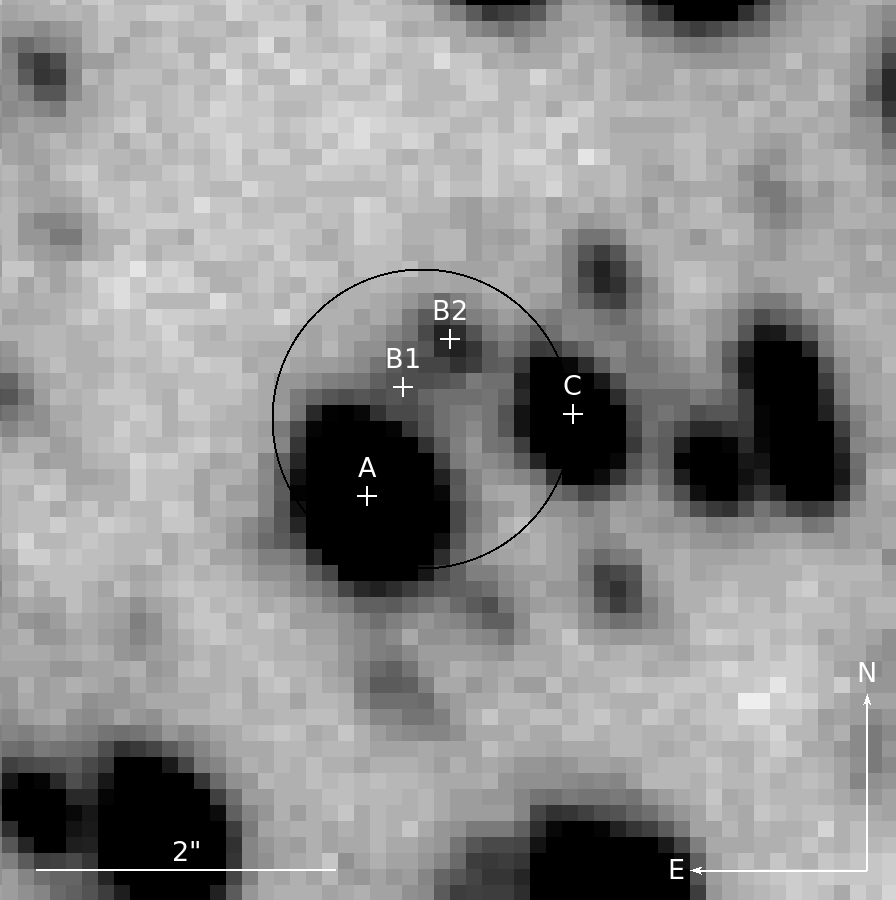}%
        \subcaption{MAXI J1543-564}
        \label{fig:j1543}
         \vspace{0.5cm}
    \end{minipage}%
	\caption{Finder charts of the two XRBs for which the outburst counterpart has not been identified until now, so we have to rely on their {\it Chandra} X-ray localisation for the identification of the counterpart in quiescence. {\it Upper panel:} LIRIS $K_s$--band image of IGR J17451-3022 in (a) outburst and (b) quiescence. As no counterparts were reported previously for IGR J17451-3022, we analysed the two sources (labeled A and B) inside the 99.7$\%$ confidence region around the {\it Chandra} X-ray position \citep{2014ATel.6533....1C}, and using differential photometry, determined that the likely counterpart is source B as that source is fainter in quiescence than in outburst. {\it Bottom panel:} HAWK-I $K_s$--band image of MAXI J1543-564, where we indicate the sources identified by \citet{2011ATel.3365....1R} and label the in the same way as they did, with the difference that they identified one source B, whereas we detect two sources in this position (B1 and B2). These two sources are consistent with the position of the likely counterpart, according to a  {\it Chandra} localization of the X-ray source \citep{2011ATel.3407....1C}.}
    \label{fig:xraypos}  
\end{figure*} 

\begin{figure}
\includegraphics[width=0.5\textwidth]{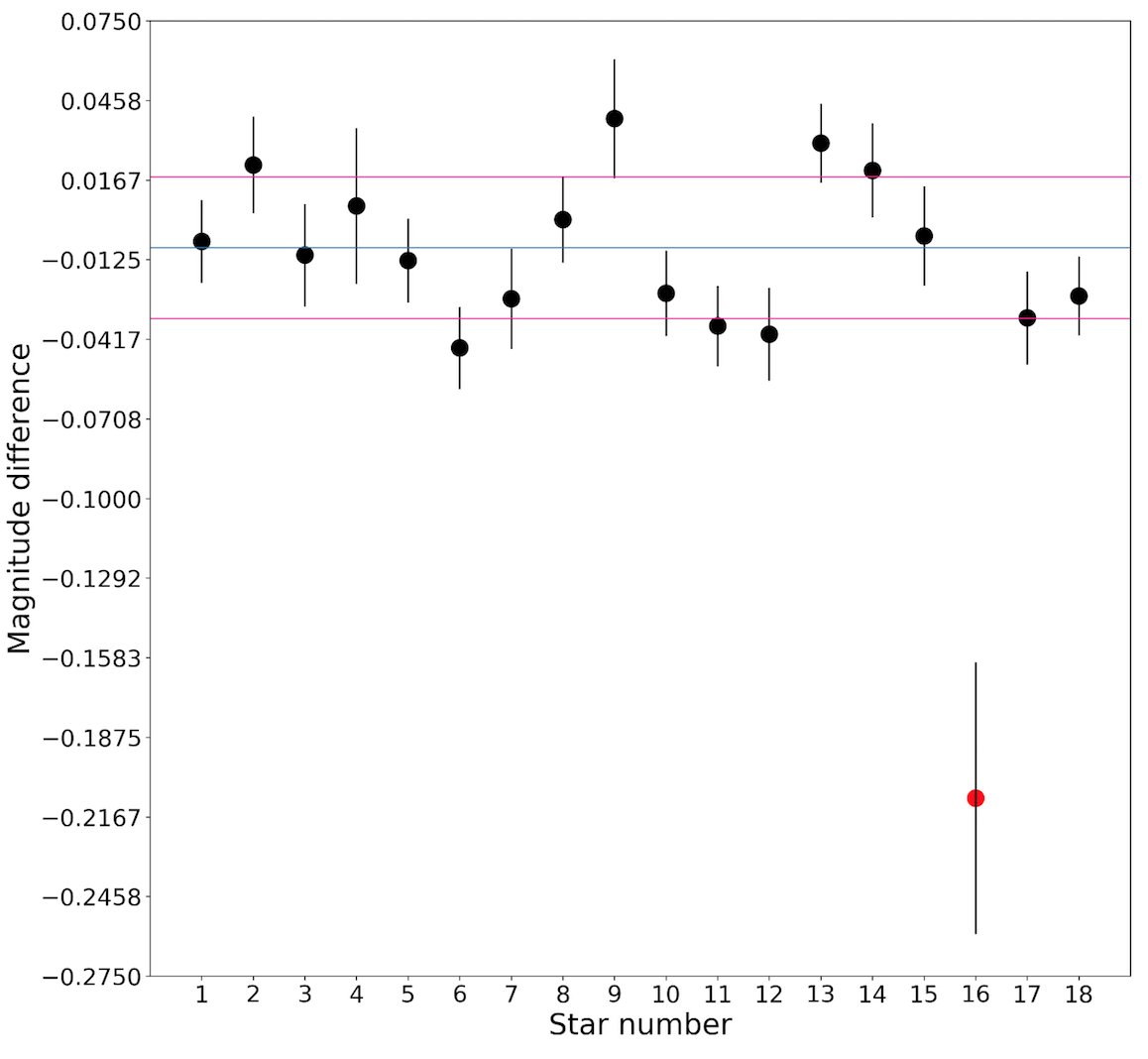}%
\caption{Magnitude difference of 18 stars inside a radius of $\sim 40$\arcsec\ around the {\it Chandra} X-ray position of IGR J17451-3022, between its outburst and quiescence images. The blue line indicates the average of all the differences and the pink line indicates the standard deviation of the distribution. Note how source B (red symbol) is clearly an outlier.}
\label{fig:variable}  
\end{figure}

\subsection{Photometry}
\label{photometry}

For seven sources (Swift J1910.2-0546, XTE J1726-476, MAXI J1807+132,
XTE J1752-223, MAXI J1543-564, Swift J1539.2-6227 and XTE J1650-500) we 
used {\scshape SExtractor} for the source detection and photometry, making sure 
that each detection was more than 3-$\sigma$ above the local background. We 
performed aperture photometry to determine instrumental magnitudes. 
We determined the full width at half maximum (FWHM) of point-like objects in each image
with the {\scshape starlink} tool {\scshape gaia} and we used it as the aperture size.
We performed point spread function (PSF) photometry for the eight 
remaining images using the {\scshape Daophot} package in {\scshape Iraf}\footnote{IRAF is distributed by the National Optical Astronomy Observatories, which are operated by the Association of Universities for Research in Astronomy, Inc., under cooperative agreement with the National Science Foundation.} because the source density warrants this.
The photometric zero points of our NIR images were measured 
by using isolated 2MASS objects (or VVV\footnote{VISTA Variables in the Via Lactea survey \citep{2010NewA...15..433M}.} objects, when possible) in the field of view,
and for our optical images we used isolated
Panoramic Survey Telescope and Rapid Response System 1 (Pan-STARRS 1, 
\citealt{2016arXiv161205243F,2016arXiv161205560C}) objects.
We calibrated the instrumental magnitudes of
those isolated sources with the magnitudes reported by their respective surveys, thus calculating the
photometric zero points (see Table~\ref{tab:sources}).  We ignore the color term in our calibrations,
as this term is small compared to other uncertainties in the photometry \citep{1992UvATAugusteijn}. 
Instrumental magnitudes for all detected sources were converted to apparent magnitudes using these zero points. 
It should be noted that the formal uncertainty in the 2MASS/Pan-STARRS 1
photometric calibration that we employ is the largest contributor ($90-95\%$) to the uncertainty
in the apparent magnitudes.
For the images in which we did not detect a counterpart, we estimated the 
limiting magnitude by simulating 10,000 stars at the position of the XRB using the
{\scshape Iraf} task {\scshape Mkobjects} and then detecting them with the task
{\scshape Daofind}. We use the faintest magnitude at which the objects were detected 
at the three sigma level as our limiting magnitude.
For the five sources (XTE J1752-223, Swift J174510.8-262411, MAXI J1836-194, MAXI J1957+032 and 
XTE J1650-500) for which we determine a limiting absolute magnitude,
we assumed $R_V = 3.1$ \citep{1999PASP..111...63F}, $R_{K_s} = 0.306$ \citep{2013MNRAS.430.2188Y} 
and $N_H = 1.87 \times 10^{21}$ Atoms cm$^{-2}$ mag$^{-1} A_V$ \citep{1978ApJ...224..132B} to 
correct for extinction. We use the distances indicated in Table~\ref{tab:coord}, and although these
estimates have large uncertainties (one value is a lower limit), even an underestimate 
of the distance by a factor of 2 would only result in a difference of 1.5 mag in absolute magnitude. 
This is the difference between spectral types K0V and K7V, or between M0V and M2V.
All the magnitudes are given in the AB system, unless indicated otherwise;
the conversion between the Vega and AB systems was done following \citet{2007AJ....133..734B}, 
i.e. $J_{AB} = J_{VEGA} + 0.91$, $H_{AB} = H_{VEGA} + 1.39$, $K_{sAB} = K_{sVEGA} + 1.85$.

\section{Results}
\label{results}

\subsection{Sources observed in outburst}
\label{outburst}

\subsubsection{IGR J17451-3022}
\label{igrj17451}

This source was discovered by INTEGRAL/JEM-X on 2014 August 22 \citep{2014ATel.6451....1C}.
Suzaku observations done in September 2014 revealed eclipses in the light curve of IGR J17451-3022 
\citep{2015ATel.7361....1J}, from which they estimated a $P_{orb}$ of $\sim$6.3 hr.
We performed a NIR observation of IGR J17451-3022 during
its outburst and detected two sources (labeled A and B in Figure~\ref{fig:igrj17451o})
in the 99.7$\%$ confidence radius around the X-ray position. The source
returned to quiescence in 2015 May \citep{2015ATel.7570....1B}. 
We observed it for a second time using the same telescope and instrument and filter on 2016 March 28 and 
detected again the same two sources (see Figure~\ref{fig:igrj17451q}) in the 99.7$\%$ confidence 
radius around the X-ray position \citep{2014ATel.6533....1C}. As is visible in 
Figures~\ref{fig:igrj17451o} and~\ref{fig:igrj17451q}, these two sources are close to 
a bright star (2MASS J17450681-3022456, $K_s$ = 13.16). To estimate the 
apparent magnitudes of the fainter sources minimizing contamination of this bright star, we subtracted it and then
performed PSF photometry on the image. Then, using differential
photometry we found that the apparent magnitude of the source labeled B varied between outburst 
and quiescence at the level of 7.8-$\sigma$ (see Figure~\ref{fig:variable}). 

The outburst and 
quiescent magnitudes are $K_s = 17.39 \pm 0.16$ and $K_s = 17.59 \pm 0.14$, respectively (see Table~\ref{tab:mags}).
Whereas these values are the same within the errors, the formal uncertainty in the 2MASS
photometric calibration that we employ is the largest contributor to the uncertainty (0.13 mag in both cases). Our
differential photometry where we calibrate one frame against the magnitudes of the second, shows
that the source in quiescence is significantly fainter by 7.8-$\sigma$.
Given that the source is eclipsing \citep{2015ATel.7361....1J},
the projected surface area of the accretion disc is
small, possibly resulting in a similar magnitude for any combination of disc 
plus star during outburst and quiescence.

\begin{figure*}
    \begin{minipage}{0.4\textwidth}
        \includegraphics[width=\textwidth]{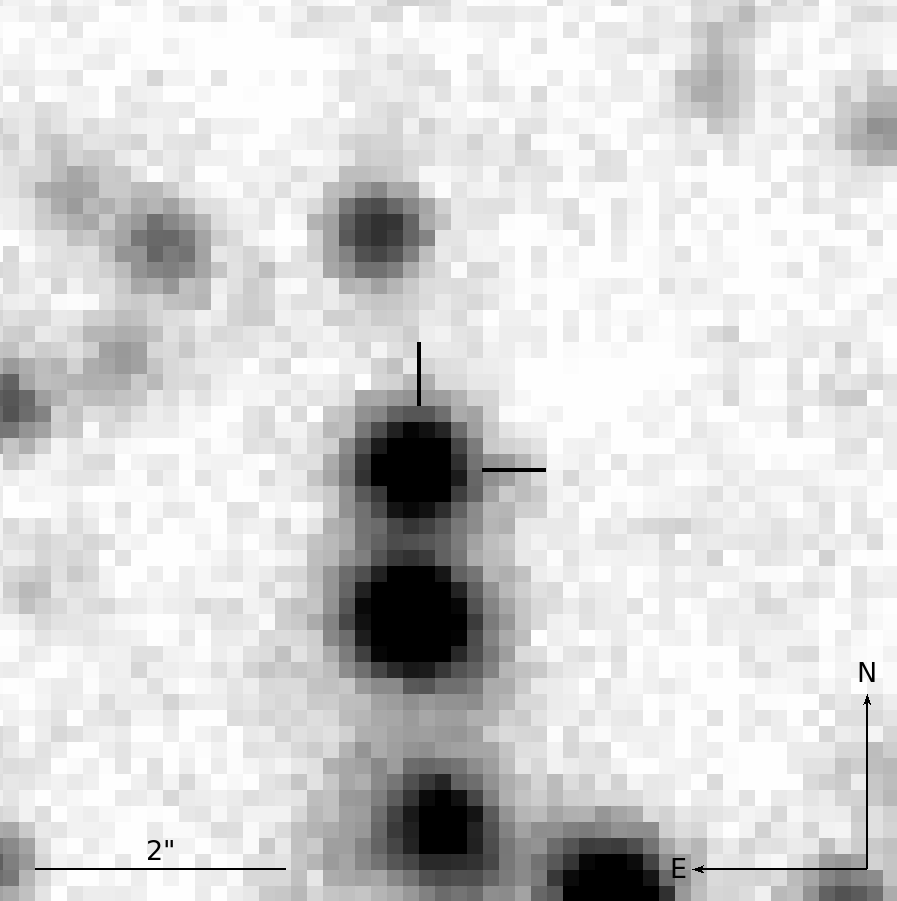}%
        \subcaption{}
        \label{fig:xtej1818o}
         \vspace{0.5cm}
    \end{minipage}%
    \begin{minipage}{0.4\textwidth}
        \includegraphics[width=\textwidth]{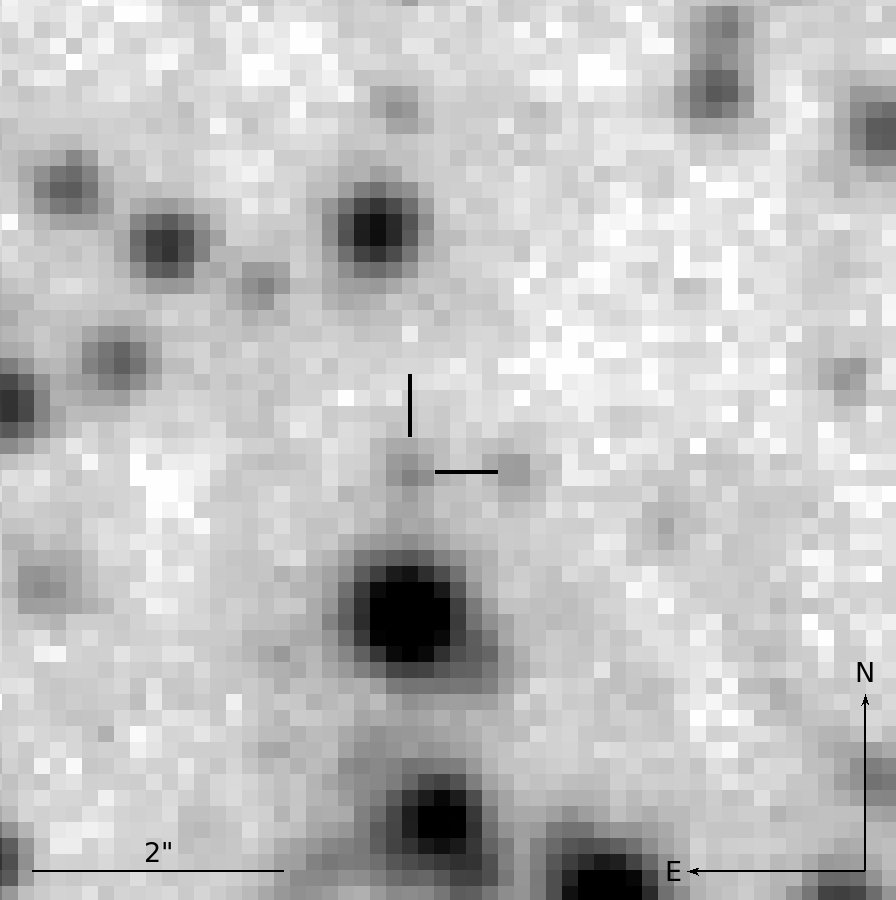}%
       \subcaption{} 
       \label{fig:xtej1818q}
        \vspace{0.5cm}       
    \end{minipage}       
	\caption{PANIC $K_s$--band image of XTE J1818-245 in (a) outburst and (b) quiescence.}
    \label{fig:outburst}  
\end{figure*} 

\subsubsection{XTE J1818-245}
\label{xtej1818}

The X-ray transient XTE J1818-245 was discovered on 2005 August 12 \citep{2005ATel..578....1L}
with the RXTE All-Sky Monitor.
\citet{2005ATel..585....1S} observed it a few days after the outburst and detected an
optical candidate counterpart with $R \sim 17.71$ mag which was later
confirmed by \citet{2011MNRAS.413..235Z}.
We observed this source a month after the outburst discovery and detected a NIR counterpart with
an apparent magnitude of $K_s = 16.18 \pm 0.02$; this is consistent with the result from
\citet{2009AA...501....1C}. They observed XTE J1818-245 in the NIR two months after the
outburst start and detected a counterpart with apparent magnitude $K = 16.9 \pm 0.2$.
We observed this source also during quiescence, approximately 10 years after the outburst ended, 
and we detected the NIR counterpart as well, with $K_s = 19.99 \pm 0.22$ (see Table~\ref{tab:mags}).
As can be seen from Figures~\ref{fig:xtej1818q} and~\ref{fig:xtej1818o}, it is
evident that the NIR source is bright during outburst and fades during quiescence.
Hence, this is the quiescent NIR ($K_s$--band) counterpart to XTE J1818-245.

\subsection{Sources observed in quiescence}
\label{quiescence}

\subsubsection{Swift J1539.2-6227}
\label{swft1539}

Swift J1539.2-6227 was discovered on 2008 November 24 by Swift/BAT \citep{2008ATel.1855....1K}. UV/optical observations made a month later revealed
a counterpart with magnitudes of $uvw2 = 18.07 \pm 0.03$ and $uvm2 = 17.96 \pm 0.04$ \citep{2009ATel.1893....1K}, which
continued until at least until 2009 March \citep{2009ATel.1958....1T}. The optical spectrum of the counterpart
revealed no Balmer lines, nor evidence for HeII 4686 \AA\, or Bowen blend emission \citep{2009ATel.1958....1T}. 
We observed this source with GROND and did not detect
the counterpart down to limiting magnitudes of $g' > 19.6$, $r' > 20.1$, $i' > 19.8$, 
$z' > 18.9$, $J > 18.9$, $H > 16.8$ and $K_s > 17.6$ (see Table~\ref{tab:mags}).

\subsubsection{MAXI J1543-564}
\label{maxij1543}

\citet{2011ATel.3330....1N} discovered this source on 2011 May 8. UV/optical observations
with Swift revealed no counterpart to this source down to limiting magnitudes
of $V >$ 19.47 and $U >$ 20.85 \citep{2011ATel.3336....1K}, which is consistent with the large absorption column towards 
this source.
\citet{2011ATel.3359....1R} detected a possible counterpart with a magnitude of $i = 19.5$. Two days after
the outburst was reported, \citet{2011ATel.3365....1R} detected the same optical counterpart
(called source A) with GROND, and reported outburst magnitudes of
$g' > 22.8$, $r' = 20.7 \pm 0.1$, $i' = 19.4 \pm 0.1$, $z' = 18.7 \pm 0.1$, $J = 17.1 \pm 0.2$, $H = 16.8 \pm 0.2$ and 
$K_s = 17.0 \pm 0.2$. Additionally, \citet{2011ATel.3365....1R} reported the detection of two other sources, 
called source B and C, with 
magnitudes $z' \sim 20.7$ and $z' \sim 20.5$, respectively. No measurements in the $JHK_s$ bands are reported as they
are strongly affected by the blending of sources A, B and C. A {\it Chandra} localization of MAXI J1543-564 
\citep{2011ATel.3407....1C} showed that the optical source detected by 
\citet{2011ATel.3359....1R} and \citet{2011ATel.3365....1R} was not likely the counterpart but instead, 
the source identified as source B by \citet{2011ATel.3365....1R} is a better candidate counterpart. 
We observe this source with HAWK-I and detect two sources inside the 3-$\sigma$ confidence radius
around the position of source B from \citet{2011ATel.3365....1R} in both $H$ and $K_s$ bands 
(see Figure~\ref{fig:j1543}). We label the sources B1 and B2 and estimate magnitudes of 
$H = 20.5 \pm 0.2$ and $K_s = 20.68 \pm 0.16$ for B1, and $H = 21.7 \pm 0.2$ and 
$K_s = 21.8 \pm 0.2$ for B2. Further observations are required in order to
determine which one is the true counterpart (if any).

\subsubsection{XTE J1650-500}
\label{xtej1650}

This XRT was discovered on 2001 September 5 by RXTE and
optical observations revealed a counterpart with  $V = 17$ \citep{2001IAUC.7707....3C}.
A month later, with observations from ESO NTT,
NIR/optical magnitudes were derived \citep{2012AA...547A..41C}: $V = 17.1 \pm 0.2, R = 16.9 \pm 0.4, I = 16.14 \pm 0.13, 
J = 15.33 \pm 0.11, H = 15.18 \pm 0.11$ and $K_s = 15.14 \pm 0.13$ mag.
It was observed $\sim$ 10 months later (2002 Aug 10) in the optical
with the Magellan telescope, and the counterpart was detected in quiescence, with magnitudes of $R \sim$ 22 and 
$V \sim$ 24 \citep{2002ATel..104....1G}.
We observed this source with GROND and did not detect
the counterpart down to limiting magnitudes of $g' > 20.3$, $r' > 20.5$, $i' > 19.4$, 
$z' > 18.5$, $J > 18.4$, $H > 16.8$ and $K_s > 17.8$ (see Table~\ref{tab:mags}),
resulting in an outburst amplitude of $>$2 mag in the $JHK_s$ bands. With a distance 
of $d = 2.6 \pm 0.7$ kpc given by \citet{2006MNRAS.366..235H}  and 
$N_H = 0.5 \pm 0.1 \times 10^{22}$ cm$^{-2}$ \citep{2004MNRAS.351..466M}, we calculate an extinction of 
$A_{K_s} = 0.270$, which gives a limiting absolute magnitude of
$M_K >5.4$, making the donor star an M0V spectral type or later.

\subsubsection{XTE J1726-476}
\label{xtej1726}

XTE J1726-476 was discovered with RXTE by \citep{2005ATel..623....1L} on 2005 October 4. 
Optical, $I$--band observations, revealed a counterpart with $I$ = 17.42 $\pm$ 0.11 mag, 
days after the outburst start \citep{2005ATel..628....1M}. 
Additionally, \citet{2005ATel..629....1S} observed XTE J1726-476 in the NIR and detected the
counterpart with $K_s = 18.05$. We observed
this source in quiescence in both $J$ and $K_s$ bands (see Table~\ref{tab:mags}). 
Our observation in the $K_s$--band yielded a non-detection down to a
limiting magnitude of $K_s > 17.9$, 
whereas in the $J$--band we just detect a faint candidate counterpart with $J = 21.0 \pm 0.3$ 
(see Figure~\ref{fig:1726}).

\subsubsection{Swift J174510.8-262411}
\label{swft1745}

This XRT went into outburst on 2012 September 16 \citep{2012GCN..13775...1C} and was observed with GROND 
one day after \citep{2012ATel.4380....1R}, where they found a candidate counterpart with an
apparent magnitude of $i' \sim 17.8$ and $J \sim 16.5 \pm 0.5$, which was then spectroscopically
confirmed by \citet{2012ATel.4388....1D}.
Subsequent optical observations were performed by \citet{2012ATel.4417....1H}
and \citet{2012ATel.4456....1R}; the former did not detect the counterpart, 
whereas the latter detected the source previously identified by \citet{2012ATel.4380....1R}, 
this time with a magnitude of $i' = 17.59 \pm 0.07$.
We observed this source 5 years after its outburst and detect the counterpart in the NIR
(see Figure~\ref{fig:1745}) with a magnitude of $K_s = 18.63 \pm 0.15$ (see Table~\ref{tab:mags}) 
with a position consistent with that reported in \citet{2012ATel.4380....1R}.
\citet{2013MNRAS.432.1133M} estimated a distance of $d > 7$ kpc (see Table~\ref{tab:coord}) 
and with $N_H = 1.70 \pm 0.04 \times 10^{22}$ cm$^{-2}$ \citep{2012ATel.4393....1T}, 
we get an extinction of $A_{K_s} = 0.918$, which we use to estimate a limiting absolute 
magnitude of $M_K > 3.4$.  This would make the donor star a K0V star or later
\citep{2000asqu.book..143T,2000asqu.book..381D}, consistent with the hydrogen emission lines
detected by \citet{2012ATel.4388....1D}.

\subsubsection{XTE J1752-223}
\label{xtej1752}

This source was discovered on 2009 October 23 \citep{2009ATel.2258....1M}, and later confirmed
to be a strong accreting black hole candidate
\citep{2010PASJ...62L..27N,2010MNRAS.404L..94M,2010ApJ...723.1817S,2011MNRAS.410..541C}.
An optical counterpart was identified by \citet{2009ATel.2263....1T}, later confirmed by optical spectra and a 
NIR detection \citep{2009ATel.2268....1T} with a magnitude of $K_s = 15.83 \pm 0.01$. Approximately 
9 months after the outburst, \citet{2010ATel.2775....1R} reported optical and NIR quiescent 
magnitudes for the counterpart of $V = 21.2 \pm 0.3$ and $K_s = 17.1 \pm 0.1$, respectively.
However, \citet{2012MNRAS.423.2656R} showed that the source had not yet reached the 
quiescent state at the point in time related to the observations of \citet{2010ATel.2775....1R},
as they did not detect the source 10 months later down to a limiting magnitude of
$i' >$ 24.77, making the difference between the source brightness in outburst and 
in quiescence at least 8 mag. We obtained NIR observations of this source 
six years after its outburst and did not detect any source at the position of the outburst counterpart, 
down to a limiting magnitude of $K_s >$ 19.6.
This is 2.5 mag fainter than reported before by \citet{2010ATel.2775....1R}.
With a distance of $d = 6 \pm 2$ kpc, derived by \citet{2012MNRAS.423.2656R} 
(see Table~\ref{tab:coord}), and using $N_H = 4.5 \times 10^{21}$ cm$^{-2}$ 
\citep{1990ARA&A..28..215D}, we get an extinction of $A_{K_s} = 0.243$, which results in a 
limiting absolute magnitude of $M_K > 5.3$, which means that the donor star is a main 
sequence star with spectral type M0 or later \citep{2000asqu.book..143T,2000asqu.book..381D}. 
This is consistent with the constraints on the spectral type given in \citet{2012MNRAS.423.2656R} and with
the hydrogen emission lines found in the spectra by \citet{2009ATel.2268....1T}.

\subsubsection{MAXI J1807+132}
\label{maxij1807}

MAXI J1807+132 was discovered on 2017 March 13 by MAXI/GSC \citep{2017ATel.10208...1N}.
A UV/optical counterpart was detected soon after by \citet{2017ATel.10216...1K} and 
\citet{2017ATel.10217...1D}, who reported magnitudes of $B = 18.3 \pm 0.2$ and $V = 17.82$, 
respectively. Furthermore, \citet{2017ATel.10217...1D} reported the quiescent magnitude of this optical
counterpart as $r' = 21.19 \pm 0.09$ and $i' = 21.38 \pm 0.04$ from PanSTARRS-1 archival images. 
Spectroscopy taken  with GTC/OSIRIS revealed spectral features consistent with that of an LMXB
\citep{2017ATel.10221...1M,2017ATel.10222...1S}.
\citet{2017ATel.10223...1T}
detected in the optical a decay in magnitude of $\sim$ 0.4 mag/day, which coincided with a decay in the X-ray
\citep{2017ATel.10224...1A}. Furthermore, after the X-ray fading
phase of MAXI J1807+132, the optical counterpart varied its magnitude from $r' = 20.76 \pm 0.07$ to
$r' = 19.34 \pm 0.02$ in a 5-day interval \citep{2017ATel.10245...1K}.
We observed this X-ray source in the $i'$--band (see Figure~\ref{fig:1807}) 
four months after its ourburst started, and detected the optical counterpart
reported by \citet{2017ATel.10216...1K} with an magnitude of $i' = 20.26 \pm 0.05$. This is 2.3 mag brighter
than the archival magnitude from PanSTARRS-1, which implies that the source had not fully
returned to quiescence yet at the time of our observations.

\subsubsection{MAXI J1828-249}
\label{maxij1828}

\citet{2013ATel.5474....1N} discovered MAXI J1828-249 in a soft state, on 2013 October 15.
An uncatalogued UV point source was detected with Swift as a possible counterpart, with a magnitude of 
$uvm2 = 18.64 \pm 0.04 \pm 0.03$ (with 0.04 and 0.03 being statistical and systematic 
uncertainties, respectively, \citealt{2013ATel.5479....1K}).
Follow-up observations by \citet{2013ATel.5482....1R} detected a
counterpart at $H = 16.9 \pm 0.1$ and $K_s = 17.2 \pm 0.2$. A
month later, another NIR observation \citep{2013ATel.5559....1D} revealed fading of this counterpart, 
with a magnitude of $H = 18.7 \pm 0.1$, which implies that this source is the counterpart.
We detect this source in our NIR observations with 
$K_s = 20.82 \pm 0.09$ (see Table~\ref{tab:mags}), 3.6 mag 
fainter than in outburst (see Figure~\ref{fig:1828}).

\begin{table*}
\vspace{5mm}
\begin{center}
\caption{Outburst and quiescent magnitudes.}
\label{tab:mags}
\resizebox{\textwidth}{!}{\begin{tabular}{l|ccc|cc}
\toprule
& \multicolumn{3}{c}{Outburst} & \multicolumn{2}{c}{Quiescent} \\
\cmidrule(lr){2-4}\cmidrule(lr){5-6}
BHXB candidate & Time after & & Reference & Time after &\\
 & discovery & & & discovery &\\
\midrule
Swift J1539.2-6227 & 1 month & $uvw2$ = 18.07 $\pm$ 0.03 & \citet{2009ATel.1893....1K} & 10.8 years &  $g' > 19.6$\\ 
 & & $uvm2$ = 17.96 $\pm$ 0.04 & &  &  $r' > 20.1$\\
 & &  & & &  $i' > 19.8$\\ 
 & &  & & &  $z' > 18.9$\\ 
 & &  & & &  $J > 18.9$\\
 & &  & & &  $H > 16.8$\\  
 & &  & & &  $K_s > 17.6$\\ 
 MAXI J1543-564 & 12 days & $z' \sim 20.7$ & \citet{2011ATel.3365....1R}  & 7 years & $H$ = 20.5 $\pm$ 0.2\\
& &  & & & $K_s = 20.68 \pm 0.16$\\ 
& &  & & & $H = 21.7 \pm 0.2$\\  
& &  & & & $K_s = 21.8 \pm 0.2$\\
XTE J1650-500 & 1 month & $V$ = 17.1 $\pm$ 0.2 & \citet{2012AA...547A..41C} & 16 years &  $g' > 20.3$\\
& & $R$ = 16.9 $\pm$ 0.4 & & & $r' > 20.5$\\
 & & $I$ = 16.14 $\pm$ 0.13 & & &  $i' > 19.4$\\
  & &  & & &  $z' > 18.5$\\
 &  & $J$ = 15.33 $\pm$ 0.11 & & & $J > 18.4$\\
 & & $H$ = 15.18 $\pm$ 0.11 & & &  $H > 16.8$\\
 & & $K_s$ = 15.14 $\pm$ 0.13 & & &  $K_s > 17.8$\\ 
XTE J1726-476 & 6 days & $I$ = 17.42 $\pm$ 0.11 &  \citet{2005ATel..628....1M} & 10 months & $J$ = 21.0 $\pm$ 0.3\\
 & 8 days & $K_s$ = 18.05 & \citet{2005ATel..629....1S}  & 10 months & $K_s$ $>$ 17.9\\
IGR J17451-3022 & 8 months & $K_s$ = 17.39 $\pm$ 0.16 & This work & 1.6 years & $K_s$ = 17.59 $\pm$ 0.14\\
Swift J174510.8-262411 & 1 day & $J \sim$ 16.5 $\pm$ 0.5 & \citet{2012ATel.4380....1R}  & 4.5 years & $K_s$ = 18.63 $\pm$ 0.15\\
XTE J1752-223 & 3 days  & $K_s$ = 15.83 $\pm$ 0.01 & \citet{2009ATel.2268....1T} & 6.5 years & $K_s$ $>$  19.6\\
MAXI J1807+132 & 14 days & $B$ = 18.3 $\pm$ 0.2 &  \citet{2017ATel.10216...1K} & -- & $i'$ = 21.38 $\pm$ 0.04\\  
XTE J1818-245 & 1 month & $K_s$ = 16.18 $\pm$ 0.02 & This work & 8 months & $K_s$ = 19.99 $\pm$ 0.22\\
MAXI J1828-249 &  3 days & $K_s$ = 17.2 $\pm$ 0.2 & \citet{2013ATel.5482....1R}  & 3.6 years & $K_s$ = 20.82 $\pm$ 0.09\\
MAXI J1836-194 & 1 day & $K_s$ = 14.00 $\pm$ 0.07 & \citet{2011ATel.3619....1R}  & 6 years & $K_s$ = 20.9 $\pm$ 0.3\\
XTE J1856+053 &  17 days$^*$ & $K_s$ = 18.28 $\pm$ 0.05 & \citet{2007ATel.1072....1T}  & 10 years$^*$ & $K_s$ = 20.09 $\pm$ 0.26\\
Swift J1910.2-0546 & 1 day & $r'$ =15.7 $\pm$ 0.1 & \citet{2012ATel.4144....1R} & 3 years & $r'$ = 23.46 $\pm$ 0.07\\
 & & $i'$ =15.6 $\pm$ 0.1 & & & $i'$ =  22.18 $\pm$ 0.04\\
 & &$K_s$ = 15.6 $\pm$ 0.1 & & 5 years & $K_s$ = 20.43 $\pm$ 0.11\\
MAXI J1957+032 & 4 days & $J$ = 19.75 $\pm$ 0.14 & \citet{2015ATel.7524....1R} & 2 years & $K_s$ = 22.29 $\pm$ 0.15\\
&  & $K_s >$ 19.0& & & \\
XTE J2012+381 & 8 days & $K$ = 16.2 $\pm$ 0.1 & \citet{1998IAUC.6933....2C} & 9 years & $K_s$ = 20.08 $\pm$ 0.22\\
 & 5 days &  $i'$ = 17.21 $\pm$ 0.04 & \citet{2012ATel.4456....1R}  & & \\
\bottomrule
\multicolumn{6}{l}{{\bf Notes: } $^*$Time after the second outburst was detected. All the values in quiescence are calculated in this manuscript, with the}\\
\multicolumn{6}{l}{exception of the magnitude for MAXI J1807+132, which is from the PanSTARRS 1 catalogue. The magnitudes are in the AB system,}\\
\multicolumn{6}{l}{unless indicated otherwise (conversion between the Vega and AB systems was done following \citealt{2007AJ....133..734B}, i.e.}\\
\multicolumn{6}{l}{$J_{AB} = J_{VEGA} + 0.91$, $H_{AB} = H_{VEGA} + 1.39$, $K_{sAB} = K_{sVEGA} + 1.85$).}\\
\end{tabular}}
\end{center}
\end{table*}

\begin{table*}
\vspace{5mm}
\begin{center}
\caption{Magnitudes previously reported as quiescent compared to our more recent (and fainter) values}
\label{tab:mags2}
\resizebox{\textwidth}{!}{\begin{tabular}{l|ccc|cc}
\toprule
& \multicolumn{3}{c}{Previous value} & \multicolumn{2}{c}{Our values} \\
\cmidrule(lr){2-4}\cmidrule(lr){5-6}
BHXB & Time after & Magnitude & Reference & Time after & Magnitude\\
candidate & discovery &  &  & discovery & \\
\midrule
XTE J1752-223$^*$ & 9 months & $K_s$ = 17.1 $\pm$ 0.1& \citet{2010ATel.2775....1R} & 6.5 years & $K_s$ $>$ 19.6\\
XTE J1856+053 & 68 days$\ddagger$ & $K_s$ = 19.7 $\pm$ 0.1 & \citet{2007ATel.1072....1T}  & 10 years$\ddagger$ & $K_s$ = 20.09 $\pm$ 0.26\\
XTE J2012+381 & 82 days & $K_s$ = 17.96 $\pm$ 0.04 & \citet{1999MNRAS.305L..49H} & 9 years & $K_s$ = 20.08 $\pm$ 0.22\\
\bottomrule
\multicolumn{6}{l}{{\bf Notes:} $^*$\citet{2012MNRAS.423.2656R} showed that the source had not yet reached the quiescent state in the $i'$--band at the}\\
\multicolumn{6}{l}{time of the \citet{2010ATel.2775....1R} observations. We here report the first measurement in the $K_s$--band. $\ddagger$Time after}\\
\multicolumn{6}{l}{the second outburst was detected.}\\
\end{tabular}}
\end{center}
\end{table*}

\subsubsection{MAXI J1836-194}
\label{maxij1836}

MAXI J1836-194 was first detected on 2011 August 29 by MAXI/GSC and Swift/BAT, as a hard X-ray transient \citep{2011ATel.3611....1N}. 
Swift observations revealed an optical counterpart 
with a magnitude of $V = 16.22 \pm 0.04$
\citep{2011ATel.3613....1K}. GROND observations detected the optical counterpart with
magnitudes of $g' = 16.21 \pm 0.05$, $r' = 15.92 \pm 0.05$, $i' = 15.53 \pm 0.01$, $z' = 15.09 \pm 0.05$, $J = 14.77 \pm 0.05$, 
$H = 14.34 \pm 0.05$ and $K_s = 14.00 \pm 0.07$ mag \citep{2011ATel.3619....1R}.
This X-ray binary has a face-on accretion disc with an inclination between 4$^{\circ}$ and 15$^{\circ}$, 
and a lower limit on the compact object mass of 1.9 $M_{\odot}$ \citep{2014MNRAS.439.1381R}.
Six years after the first discovery outburst, we just detect the NIR counterpart at the position given by \citet{2011ATel.3613....1K}, 
with a magnitude of $K_s = 20.9 \pm 0.3$ (see Figure~\ref{fig:1836}). 
This source has an estimated distance of $d = 7 \pm 3$ kpc (\citealt{2014MNRAS.439.1381R}, 
see Table~\ref{tab:coord}) and a $N_H = 2.0 \pm 0.4 \times 10^{21}$ cm$^{-2}$ 
\citep{2011ATel.3613....1K}, which results in an extinction of $A_{K_s} = 0.108$ and an absolute 
magnitude of $M_K > 6.5$. This would make the donor star a M2 main sequence star or later
\citep{2000asqu.book..143T,2000asqu.book..381D}.

\subsubsection{XTE J1856+053}
\label{xtej1856}

This BHXB candidate was discovered on 1996 September 17 \citep{1996IAUC.6504....2M}, with a new outburst
observed 9 years later on 2007 February 28 \citep{2007ATel.1024....1L}.
It seems that the source was already in outburst around 2007 February 22 \citep{2007ATel.1093....1K}.
Days after the 2007 outburst detection, NIR observations were performed and yielded a counterpart with
an apparent magnitude of $K_s = 18.28 \pm 0.05$ \citep{2007ATel.1072....1T} which had faded two
months after the outburst to $K_s = 19.7 \pm 0.1$. In 2015, a new outburst from
XTE J1856+053 was detected by MAXI/GSC \citep{2015ATel.7233....1S,2015ATel.7579....1N} 
and Swift/XRT \citep{2015ATel.7278....1S}. Two years later, we observed this source in the NIR (see Figure~\ref{fig:1856}). 
We detect the same counterpart reported by \citet{2007ATel.1072....1T}, this time with a magnitude of
$K_s = 20.09 \pm 0.26$, which is consistent at the 1-$\sigma$ level with the quiescent 
value reported in 2007 (see Table~\ref{tab:mags2}).

\subsubsection{Swift J1910.2-0546}
\label{swft1910}

An outburst from Swift 1910.2-0546 was discovered on 2012 May 31
 \citep{2012ATel.4140....1U}. 
Optical/NIR observations made one day after by \citet{2012ATel.4144....1R} revealed a counterpart
with $g' = 16.1 \pm 0.1$, $r' =15.7 \pm 0.1$, $i' =15.6 \pm 0.1$, 
$z' = 15.3 \pm 0.1$, $J = 15.5 \pm 0.1$, $H = 15.5 \pm 0.1$ and $K_s = 15.6 \pm 0.1$ mag. 
That same day, \citet{2012ATel.4146....1C} also detected this counterpart at $R = 16.11$. 
The optical spectra reported by \citet{2012ATel.4210....1C} are typical of an LMXB.
In our $K_s$--band observations (see Figure~\ref{fig:1910}), we detect the quiescent counterpart with
an apparent magnitude of $K_s = 20.43 \pm 0.11$, while in our optical observations the counterpart
is detected with an apparent magnitude of $r' = 23.46 \pm0.07$ and $i' = 22.18 \pm 0.04$.

\subsubsection{MAXI J1957+032}
\label{maxij1957}

Discovered on 2015 May 11 \citep{2015ATel.7504....1N,2015ATel.7506....1C}, 
MAXI J1957+032 was classified as an LMXB \citep{2017ATel.10221...1M}. 
Subsequent optical and NIR observations
revealed a candidate counterpart with magnitudes $r' = 20.03 \pm 0.14$ and
$J = 19.75 \pm 0.14$ \citep{2015ATel.7524....1R}. 
In the $K_s$--band there was no detection, down to a limiting magnitude
of $K_s >$ 19.0. A day later the source had faded to $r' = 21.36 \pm 0.14$, and since
this quick decay was similar to the one observed in X-rays \citep{2015ATel.7520....1M}, it is most likely 
that the NIR source is the counterpart. This source seemed to have brightened approximately 1 magnitude
when observed in the optical some months later \citep{2015ATel.8149....1G}, coinciding with a new outburst
from MAXI J1957+032 \citep{2015ATel.8143....1S}. Optical spectra were taken \citep{2016ATel.9649....1B}, which
turned out to be featureless. This is consistent with a high-inclination
compact system, where the optical spectra have almost undetectable emission lines \citep{2016A&A...587A.102B}.
MAXI J1957+032 shows short duration outbursts ($<$ 5 days) and it has a high recurrence rate (4 in 16 months).
\citet{2017MNRAS.468..564M} found not only that the optical spectra is consistent with 
MAXI J1957+032 being a short-period LMXB, but also that the short duration and short recurrence
time of the outburst are reminiscent of an accreting milisecond X-ray pulsar.
We detect an NIR source (see Figure~\ref{fig:1957}), at the position 
indicated by \citet{2015ATel.7524....1R}, with $K_s = 22.29 \pm 0.15$ mag.
Using the distance estimate of $d \sim 6$ kpc and the column density 
$N_H = 1.7 \times 10^{21}$ cm$^{-2}$ from 
\citet{2017MNRAS.468..564M} (see Table~\ref{tab:coord}), we get an extinction of $A_{K_s} = 0.091$, 
and thus, an absolute magnitude of $M_K > 8.3$, making it consistent with an M6
main sequence star or later \citep{2000asqu.book..143T,2000asqu.book..381D}.

\subsubsection{XTE J2012+381}
\label{xtej2012}

This source went into outburst on 1998 May 24 \citep{1998IAUC.6920....1R}.
A radio source was identified as a possible counterpart \citep{2000arxt.confE..88H},
along with an optical source ($V = 17.22 \pm 0.2$ mag, \citealt{1998IAUC.6932....2H}) and a NIR source 
($K$ = 16.2 $\pm$ 0.1 mag, \citealt{1998IAUC.6933....2C}). Subsequent optical and NIR observations
(done on 1998 August 14) revealed that
the NIR source faded by 1.8 mag, from $K = 16.2 \pm 0.1$ to $K_s = 17.96 \pm 0.04$ 
\citep{1999MNRAS.305L..49H} in 71 days. Moreover, this source was identified as the optical counterpart to the
radio source identified by \citet{2000arxt.confE..88H}.
In our NIR observations (see Figure~\ref{fig:2012}), performed 8 years after the outburst, we 
detect the same source, with a 
magnitude of $K_s = 20.08 \pm 0.22$. We report this as the quiescent magnitude of the counterpart (see Table~\ref{tab:mags2}).

\section{Discussion}
\label{discussion}

\begin{figure*}
    \begin{minipage}{0.46\textwidth}
        \includegraphics[width=\textwidth]{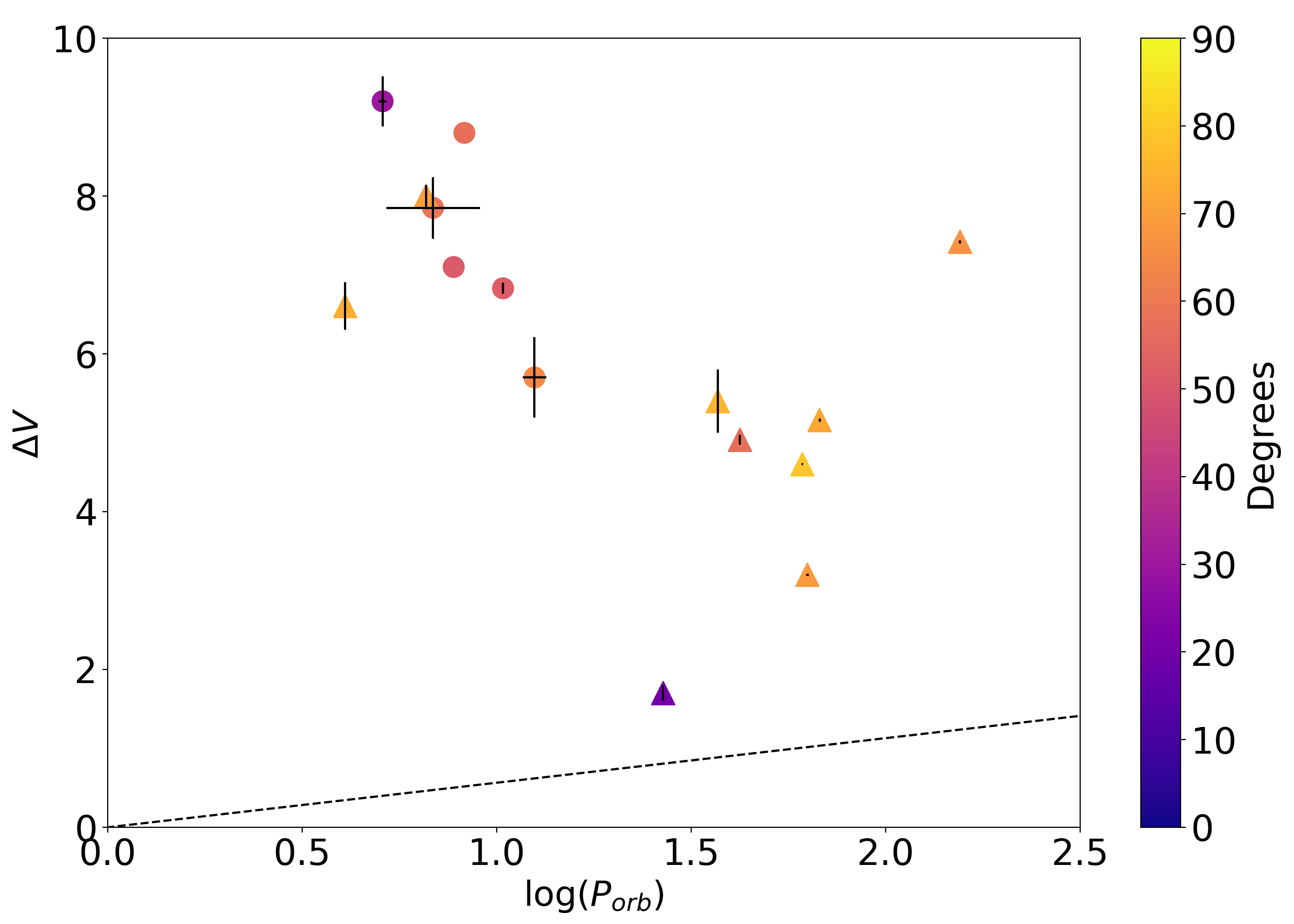}%
        \subcaption{}
        \label{fig:3dV}
         \vspace{0.5cm}
    \end{minipage}\hspace{5mm}%
    \begin{minipage}{0.46\textwidth}
        \includegraphics[width=\textwidth]{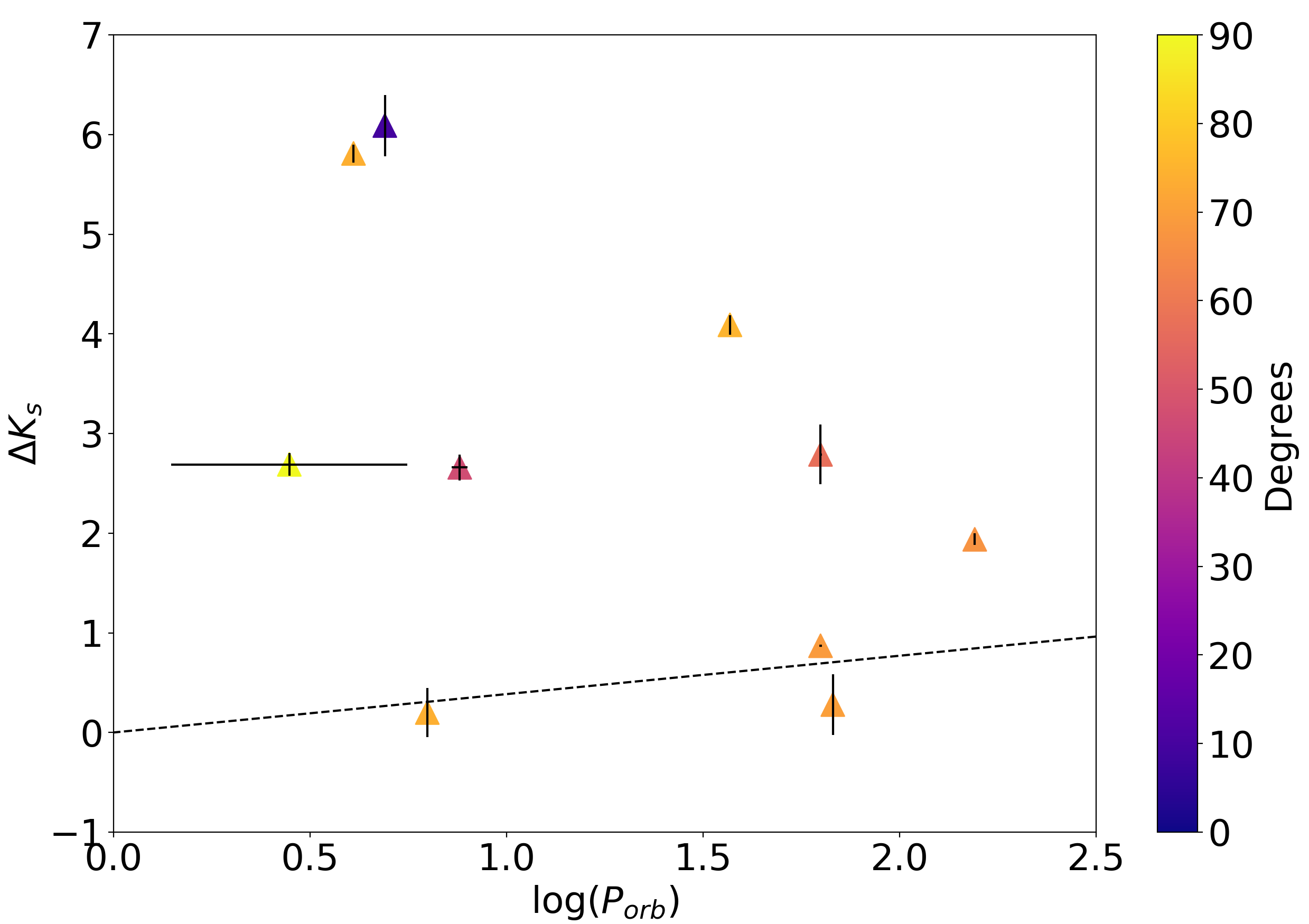}%
       \subcaption{} 
       \label{fig:3dK}
        \vspace{0.5cm}       
    \end{minipage}        
	\caption{(a) Plot of the $V$--band outburst amplitude versus the logarithm of the orbital period $P_{orb}$ and the inclination angle $i$, where the latter is indicated with different colors (for references see Table~\ref{tab:lit}); adapted from \citet{1998MNRAS.295L...1S}. The sources for which \citet{1998MNRAS.295L...1S} found a relation are indicated in round symbols, while the sources taken from other publications are indicated in triangular symbols. (b) Same as for panel (a) but now for the $K_s$--band amplitude. The dashed line indicates the relations $\Delta V \propto \log P_{orb}^{0.565}$. No error bars are shown if the original work did not provide uncertainties to their estimates.}
    \label{fig:outamp} 
\end{figure*}

\citet{1998MNRAS.295L...1S} derived an empirical relation between $\Delta V$ and $P_{orb}$ for 8 BH XRTs, 
where $P_{orb} <$ 1 day. The relation is an inverse linear correlation, 
which they suggest is explained by the fact that at longer $P_{orb}$, the secondary star is larger and hence, more luminous than a 
system with shorter $P_{orb}$, making the outburst amplitude smaller. 
As already mentioned in \citet{2011ATel.3358....1M} and \citet{2012MNRAS.421.2846R}, 
the inclination $i$ is an important parameter as well. We searched the literature and found $i$ for 6 of these 8 sources (see Table~\ref{tab:lit})
and plot $\Delta V$ versus $i$ and $P_{orb}$ in Figure~\ref{fig:3dV}. 
We also include other sources
for which $\Delta V$, $i$ and $P_{orb}$ have become available since the work of 
\citet{1998MNRAS.295L...1S} (see Table~\ref{tab:lit}). Since many of the sources analysed
are known to have had multiple outbursts (e.g. XTE J1550-564, GX 339-4, GS 2023+338), we take 
the brightest outburst magnitude and the faintest quiescent magnitude reported. The latter is because
there are sources like A 0620-00 whose mean brightness varies even 
while in quiescence ("active" and "passive" states, \citealt{2008ApJ...673L.159C}). However, note 
that the outburst amplitudes are probably lower limits, to account for the fact that one may have missed 
the peak magnitude in the sparsely sampled outbursts and that there could be a brightness variation
in quiescence.

For 10 of the 15 sources that we observed we have outburst and quiescent magnitudes in the 
$K_s$--band (see Table~\ref{tab:mags}), allowing us to derive a lower limit on
the outburst amplitude $\Delta K_s$.
As with the optical data, we complement our data with that available in the literature.
We plot $\Delta K_s$ versus $i$ and $P_{orb}$ and show the results in Figure~\ref{fig:3dK}.
We performed a 3D least squares plane fit (see Figure~\ref{fig:planes}) for both data sets and obtained

\begin{equation}
\Delta V = (-2.63 \pm 1.01)\log P_{orb} + (0.03 \pm 0.03)i + (7.82 \pm 1.94)
\end{equation}

with $\chi^2 = 16.73$ and residuals plotted in Figure~\ref{fig:resv}, and 

\begin{equation}
\Delta K_s = (-1.32 \pm 0.99)\log P_{orb} - (0.04 \pm 0.03)i + (6.89 \pm 2.16)
\end{equation}

with $\chi^2 = 2.81$ and residuals plotted in Figure~\ref{fig:resk}. Taking into account the $\chi^2$ values\footnote{To calculate $\chi^2$ we assigned to the data points without reported uncertainties, the mean uncertainty of $P_{orb}$, $i$, $\Delta V$ or $\Delta K_s$ within their respective data sets.} 
and the residuals, we see that the fit is not adequate for either data set. 
The Pearson correlation coefficient between $\Delta V$ and 
$\log P_{orb}$ is $r = -0.56$ and the coefficient between $\Delta V$ and $i$ is 
$r = 0.03$, whereas the coefficient between $\Delta K_s$ and $\log P_{orb}$ is 
$r = -0.45$ and the coefficient between $\Delta K_s$ and $i$ is $r = -0.46$.
Based on these coefficients we conclude that here is no
strong correlation between the variables.

To explore these results further, we derive the approximate size of 
the projected area of the accretion disc of a system with a given mass ratio $q$, $i$,
$P_{orb}$ and a mass $M_1$ for the compact object. From geometrical arguments, we know that the 
projected area $A_p$ of the disc can be approximated by $A_p = \pi (R_{acc} \cos i)^2$ where $R_{acc}$ 
is the accretion disc radius. Following \citet{2002apa..book.....F}, we derive $R_{acc}$ as 

\begin{equation}
R_{acc} = \frac{1}{2}P_{orb}^{\frac{2}{3}}(1+q)^{\frac{4}{3}}\Big(\frac{M_1}{M_{\odot}}\Big)^{\frac{1}{3}}(0.500-0.227\log q)^{4} R_{\odot}
\end{equation}

where $P_{orb}$ is in hours. Clearly,
$A_p$ increases with $P_{orb}$ and decreases with $i$. It was shown and parametrised by
\citet{1994A&A...290..133V} that, besides $P_{orb}$ and $i$, $L_X$ is also important for the absolute
magnitude of the disc in outburst. They found that the absolute magnitude during outburst
$M_{V,O}$ depends on $P_{orb}$ and the outburst X-ray luminosity $L_{X,O}$, as 
$M_{V,O} \propto \log (P_{orb}^{2/3}L_{X,O}^{1/2})$. The outburst X-ray luminosity also depends 
on $P_{orb}$ ($L_{X,O} \propto P_{orb}^{0.64}$, \citealt{0004-637X-718-2-620}). 

Another factor is that the quiescent X-ray luminosity shows a dependence on $P_{orb}$ as well 
(i.e. \citealt{2000A&A...360..575L,2001ApJ...553L..47G,2007ApJ...665L.147J,2008NewAR..51..752L,2009MNRAS.396L..26D,2011ApJ...729L..21R,2012MNRAS.423.3308J}), 
of the form $L_{X,Q} \propto P_{orb}^{1.77}$. 
Moreover, the quiescent magnitude $M_{V,Q}$ might be the result of (a) 
the contribution from the light of disc, which, 
like in outburst, can in principle be due to reprocessing of the X-ray light on its surface \citep{1994A&A...290..133V}, 
but also due to energy generation on the disc itself, although we ignore this below; 
and (b) the light from the donor star $M_{V,donor}$, which in outburst is difficult to detect. 
In some cases the donor star is not even detected in the 
quiescent state (e.g. \citealt{2015MNRAS.450.4292T}). 
Thus, $M_{V,Q} \propto \log (P_{orb}^{2/3}L_{X,Q}^{1/2}) + M_{V,donor}$.
If we ignore irradiation of the donor star in both outburst and quiescence, filling in
all the dependencies and relations given above, we derive
$\Delta V = M_{V,Q} - M_{V,O} \propto \log P_{orb}^{0.565}$ (we plot this correlation
as a dash line in Figure~\ref{fig:outamp}).
Therefore, for a given $i$, the same $L_{X,O}$, and $q$, systems would be expected to 
have slightly larger outburst amplitudes for longer $P_{orb}$, counter to the relation reported in \citet{1998MNRAS.295L...1S}. A factor not included above is that in the NIR, emission from a non-
thermal component, like a jet (e.g. 
\citealt{2002ApJ...573L..35C,2006MNRAS.371.1334R,2007MNRAS.379.1108R,2008MNRAS.387..713R}),
possibly contributes to the observed NIR light. 

In order to try and assess the influence that the low number of sources involved in the work of 
\citet{1998MNRAS.295L...1S} (and also to a lesser extent in the work presented here), we
determine the probability of randomly selecting 8 systems and finding a correlation between 
$\Delta V$ and $P_{orb}$. We find that it ranges between 17.5$\%$ and 30.5$\%$ (based on 
selecting 1000 different samples of 8 elements from a simulation of 100 systems with a flat 
distribution between $0 < P_{orb} < 12$ and $0 < \Delta V < 10$). We conclude that the correlation 
reported by \citet{1998MNRAS.295L...1S} was probably caused by small number statistics.

We do note that some of the observational correlations
used above to determine the expected relation between $P_{orb}$, $i$ and $\Delta V$ 
($\Delta K_s$) may suffer from low number statistics themselves implying that the predicted
correlation between $\Delta V$ and $P_{orb}$ itself is uncertain.

\section{Conclusions}
\label{conclusions}

We observed 15 BHXB candidates in quiescence in the NIR and optical in order 
to investigate which of these
sources would be promising targets for BH mass determinations. We detected 2 with
$K_s < 20$ (AB magnitude) out of our sample, which depending on their $P_{orb}$,
may allow for time-resolved spectroscopic observations to be taken with 8-10 m class
telescopes. Such measurements for fainter systems will have to await the arrival of larger
or space based telescopes.
Of our 15 observed sources, XTE J1818-245 and IGR J17451-3022, 
were observed in quiescence and in outburst, while the 
other 13 were observed only in quiescence. XTE J1818-245 presents an outburst--quiescence
magnitude difference of about $\sim$ 4 mag in the $K_s$ band. IGR J17451-3022 was analysed with 
differential photometry, as it was not clear which, if any, of the two
sources inside the 99.7$\%$ confidence
radius around it is the counterpart. One source has a magnitude change consistent in direction with the
fading in the X-rays. However, the outburst -- quiescence magnitude difference is small.
It is conceivable that this is due to the fact that the 
system has a high inclination, so the projected surface area of the disc is small.
Out of the 13 sources observed only in quiescence, we detect NIR and optical counterparts
for ten of them, while for the remaining three sources we report the limiting magnitudes. 
Of the ten sources with detected quiescent counterparts, we present measurements for the first time 
for five of them, and in contrast, two of them had previously published quiescent 
magnitudes that we update now with 
fainter magnitudes, showing that these BHXBs had not yet reached their quiescent states at the epoch 
of the observations reported in the literature before.
We analysed the $K_s$--band and $V$--band outburst amplitudes with respect to $P_{orb}$ and $i$ of 
the system for sources in our sample and for
BHXBs for which we could find published values of $\Delta K_s$, $\Delta V$, $P_{orb}$ and $i$.
We find best fits to the data although the fits are not formally acceptable, showing that probably 
additional parameters to $P_{orb}$ and $i$ are important (e.g. emission from a jet, energy generation
on the disc, light contribution from the donor star). Using theoretical arguments and observed 
relations we further derive a correlation between $\Delta V$ and $P_{orb}$ of 
$\Delta V \propto \log P_{orb}^{0.565}$, however, 
such a correlation also provides no good fit to the data, probably due to the observed relations being
affected by low number statistics, which could make our derived correlation less certain, and reinforcing
the idea that additional parameters to $P_{orb}$ and $i$ should be considered.

\section*{Acknowledgements}
PGJ and KML acknowledge funding from the European Research Council under ERC Consolidator Grant agreement no 647208. 
Part of the funding for GROND (both hardware as well as personnel) was generously granted from the Leibniz-Prize to 
Prof. G. Hasinger (DFG grant HA 1850/28-1). MAPT acknowledges support via a Ram\'on y Cajal Fellowship (RYC-2015-17854). MAPT also acknowledges support by the Spanish Ministry of Economy, Industry and Competitiveness under grant AYA2017-83216-P.
This research is based on observations made with the William Herschel Telescope operated on the island of La Palma by the Isaac Newton Group in the Spanish Observatorio del Roque de los Muchachos of the Instituto de Astrof\'{i}sica de Canarias; with the Keck Telescope at the W. M. Keck Observatory, which is operated as a scientific partnership among the California Institute of Technology, the University of California and the National Aeronautics and Space Administration. The Observatory was made possible by the generous financial support of the W. M. Keck Foundation. 
The authors wish to recognize and acknowledge the very significant cultural role and reverence that the summit of Maunakea has always had within the indigenous Hawaiian community.  We are most fortunate to have the opportunity to conduct observations from this mountain.
We have made use of the SIMBAD database, operated at CDS, Strasbourg, France; of the NASA/IPAC Extragalactic Database (NED) which is operated by the Jet Propulsion Laboratory, California Institute of Technology, under contract with the National Aeronautics and Space Administration.




\bibliographystyle{mnras}
\bibliography{library}


\appendix

\section{Quiescent detections}
\begin{figure*}
    \begin{minipage}{0.333\textwidth}
       \includegraphics[width=\textwidth]{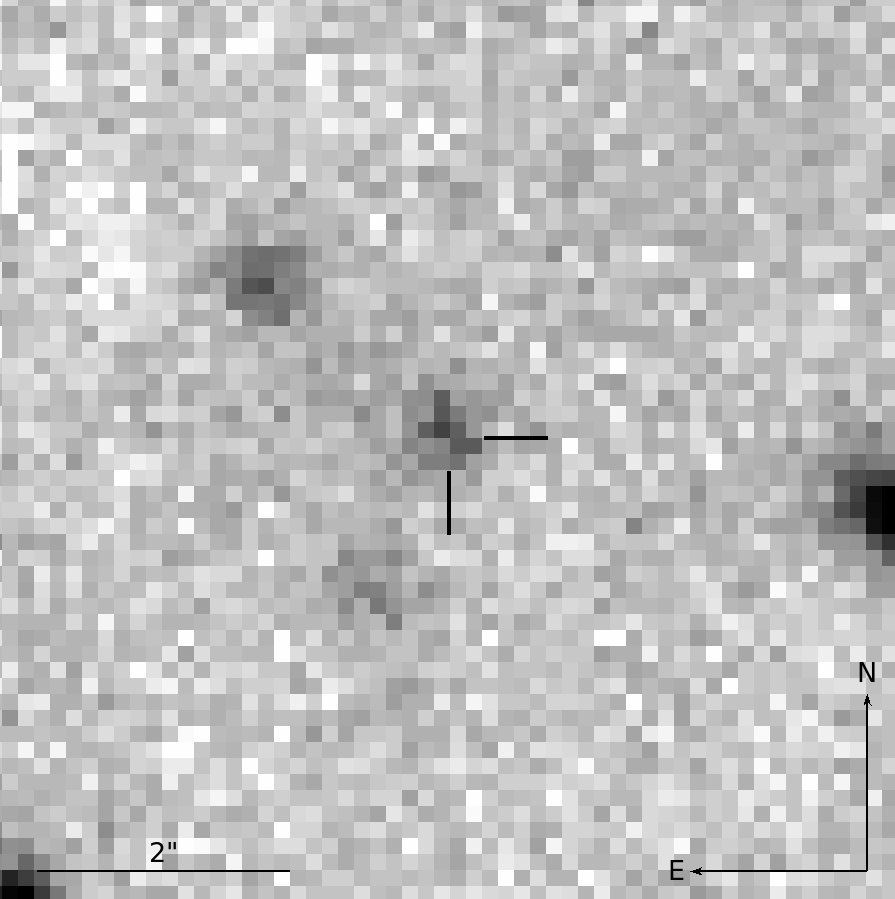}%
        \subcaption{XTE J1726-476}  
         \vspace{0.5cm}              
         \label{fig:1726} 
    \end{minipage}%
    \begin{minipage}{0.333\textwidth}
        \includegraphics[width=\textwidth]{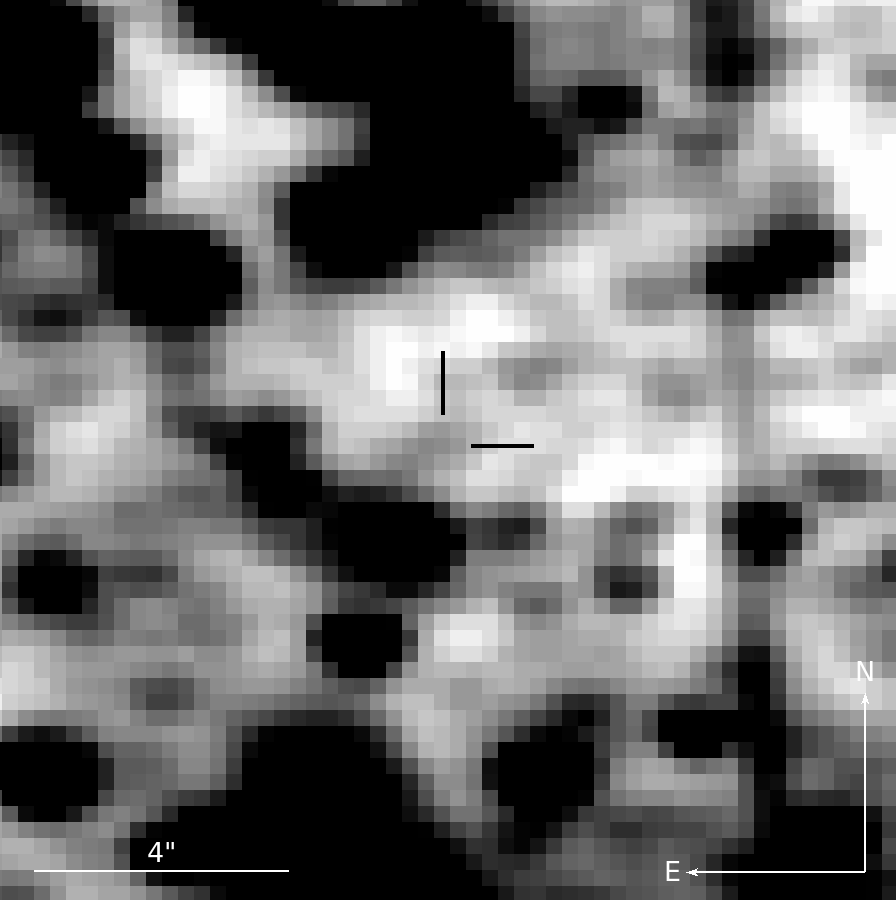}%
        \subcaption{Swift J174510.8-262411} 
         \vspace{0.5cm}               
         \label{fig:1745} 
    \end{minipage}%
    \begin{minipage}{0.333\textwidth}
        \includegraphics[width=\textwidth]{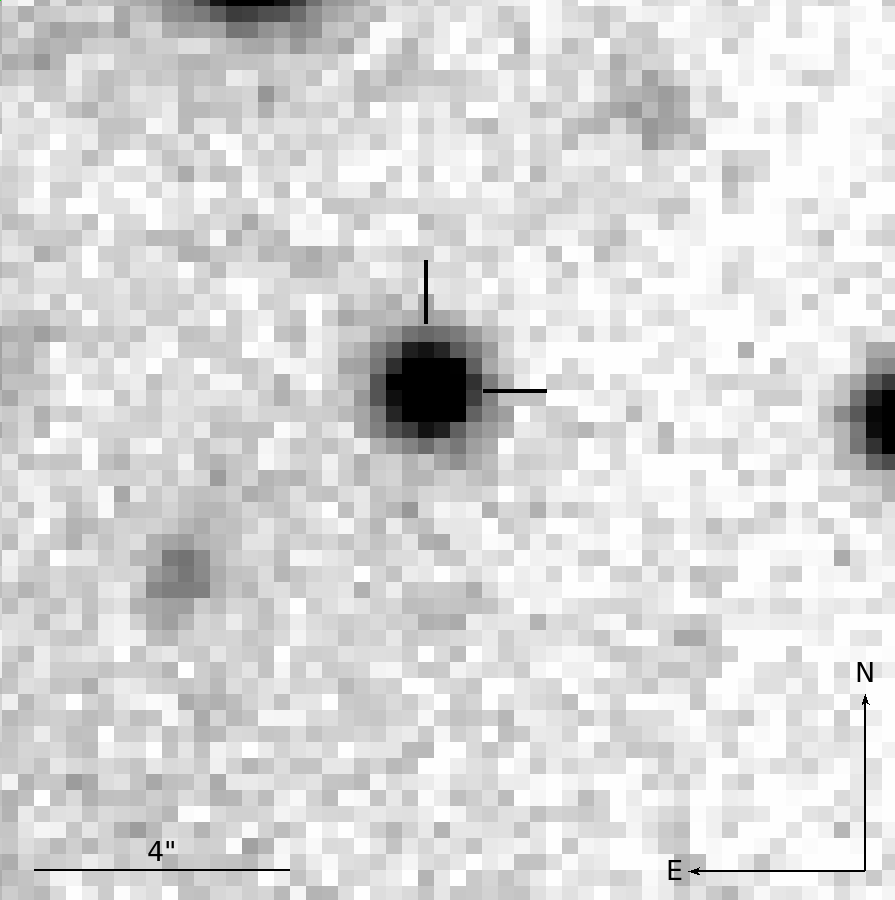}%
        \subcaption{MAXI J1807+132, $i'$--band} 
         \vspace{0.5cm}               
         \label{fig:1807} 
    \end{minipage}
        \begin{minipage}{0.333\textwidth}
        \includegraphics[width=\textwidth]{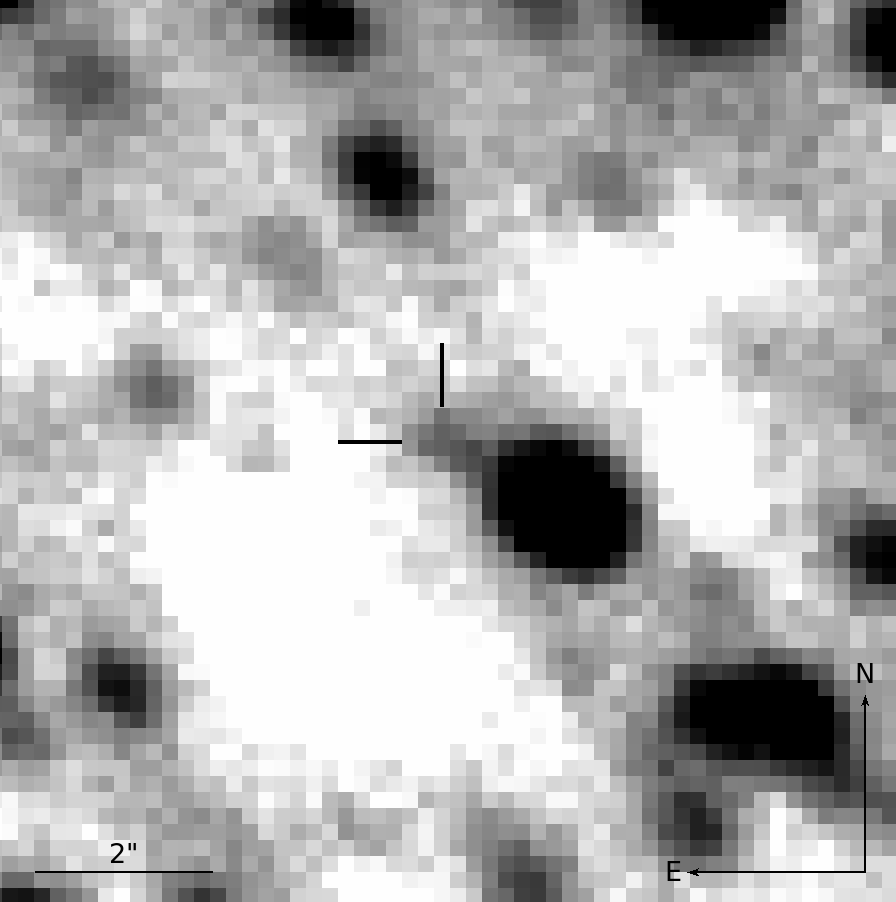}%
        \subcaption{MAXI J1828-249}
         \vspace{0.5cm}                
         \label{fig:1828} 
    \end{minipage}%
    \begin{minipage}{0.333\textwidth}
        \includegraphics[width=\textwidth]{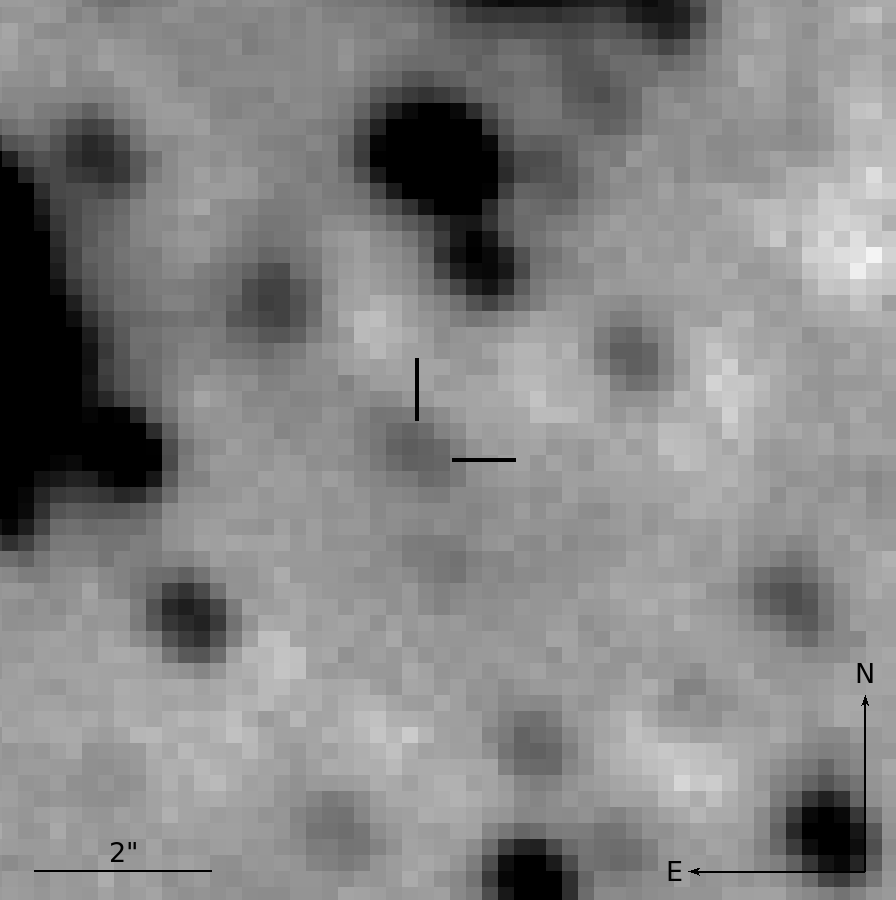}%
        \subcaption{MAXI J1836-194}
         \vspace{0.5cm}                     
         \label{fig:1836} 
    \end{minipage}%
    \begin{minipage}{0.333\textwidth}
        \includegraphics[width=\textwidth]{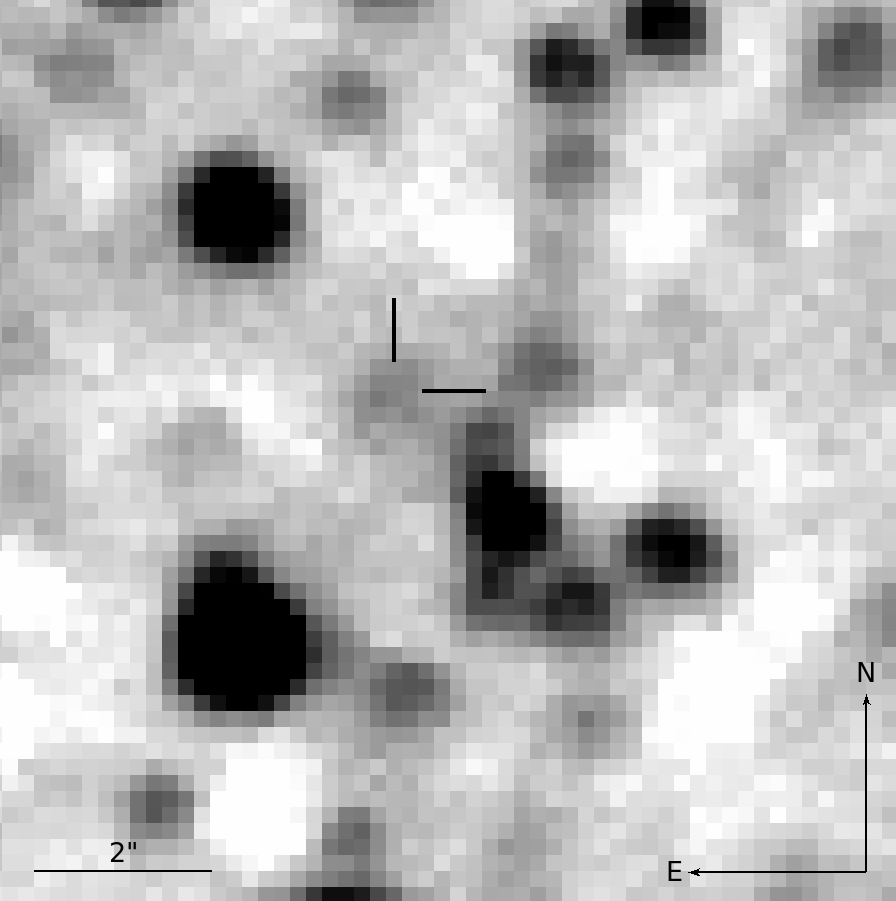}%
         \subcaption{XTE J1856+053}  
         \vspace{0.5cm}              
         \label{fig:1856} 
    \end{minipage} 
    \begin{minipage}{0.333\textwidth}
        \includegraphics[width=\textwidth]{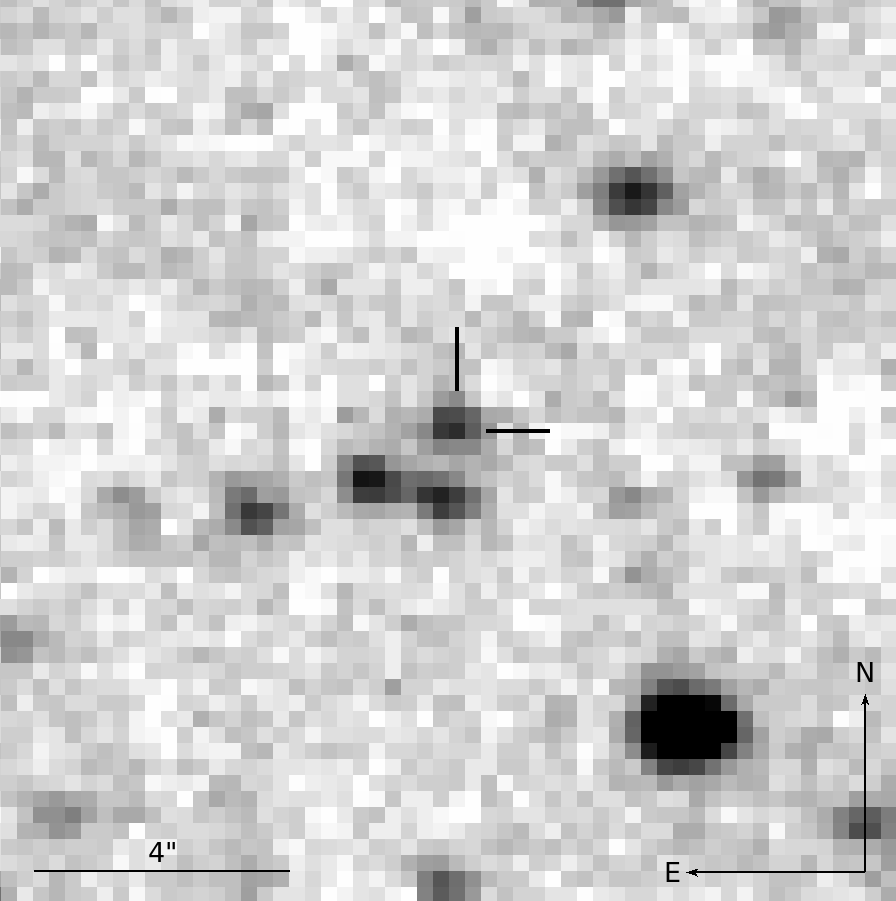}%
         \subcaption{Swift J1910.2-0546} 
         \vspace{0.5cm}
         \label{fig:1910}               
    \end{minipage}%
    \begin{minipage}{0.333\textwidth}
        \includegraphics[width=\textwidth]{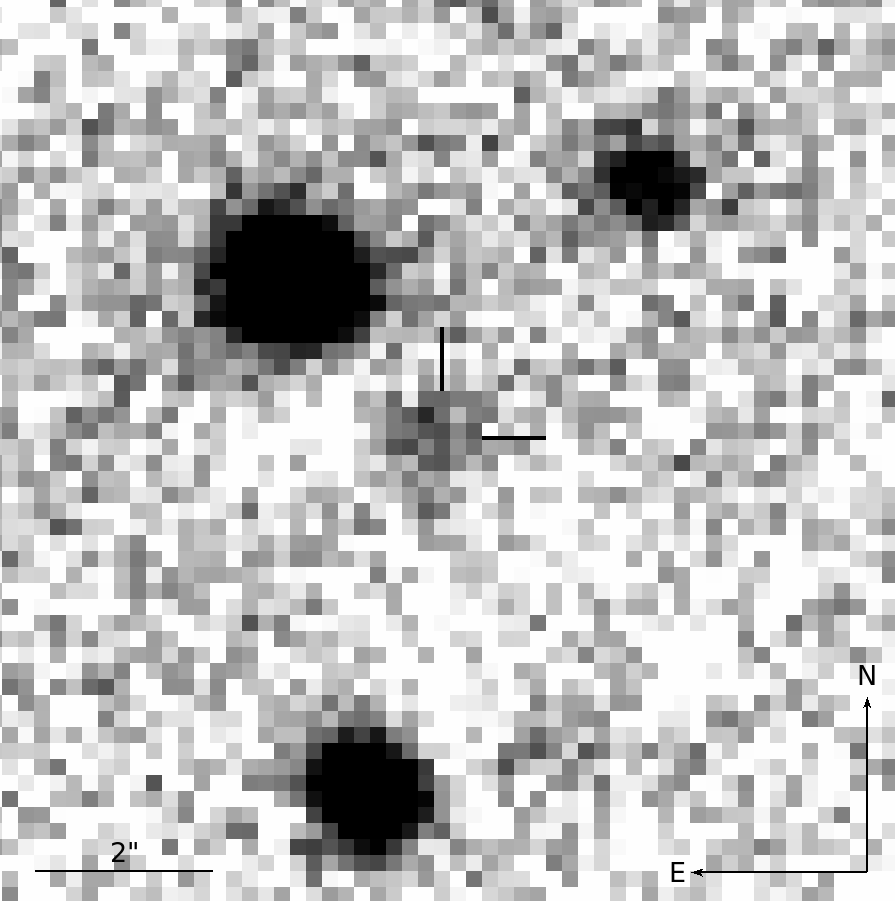}%
        \subcaption{MAXI J1957+032}  
         \vspace{0.5cm}              
         \label{fig:1957} 
    \end{minipage}%
        \begin{minipage}{0.333\textwidth}
        \includegraphics[width=\textwidth]{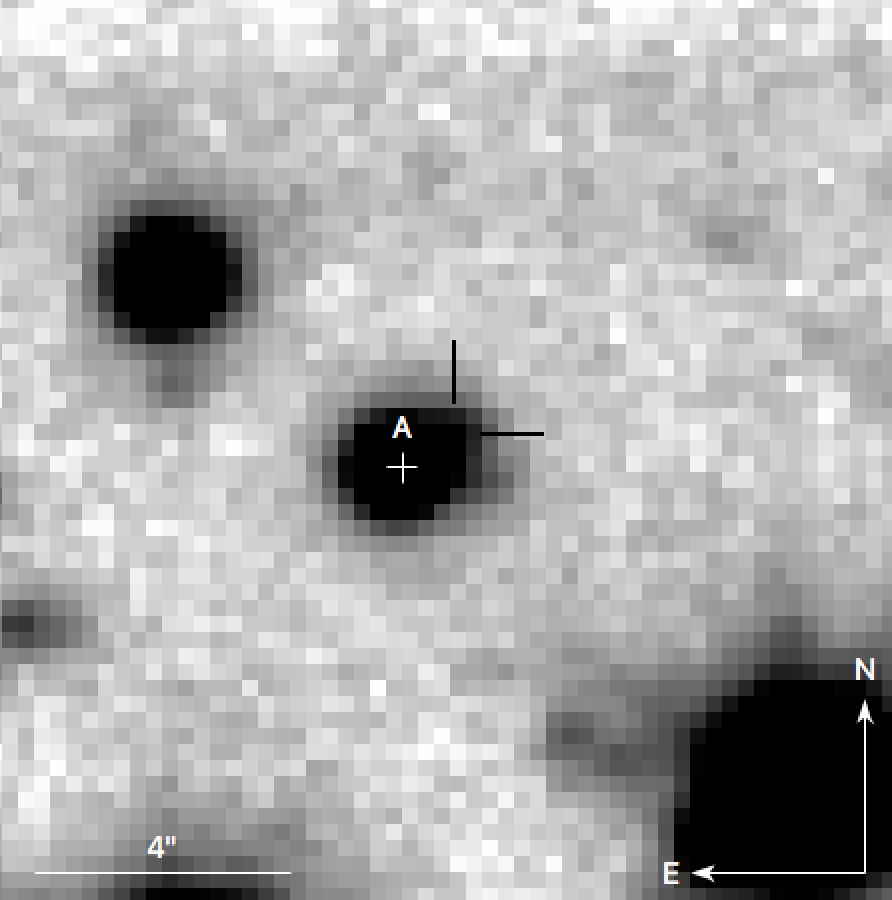}%
         \subcaption{XTE J2012+381}  
         \vspace{0.5cm}              
         \label{fig:2012} 
    \end{minipage} 
\caption{Finder charts of the XRBs with a quiescent counterpart detected. These sources have a known position for the counterpart in outburst. All images are in the NIR $K_s$--band, unless indicated otherwise. The position of the detected counterparts is indicated in every image. For image (i), we also indicate the other star identified by \citet{1999MNRAS.305L..49H} and label it A.}
\end{figure*}

\begin{figure*}
    \begin{minipage}{0.46\textwidth}
        \includegraphics[width=\textwidth]{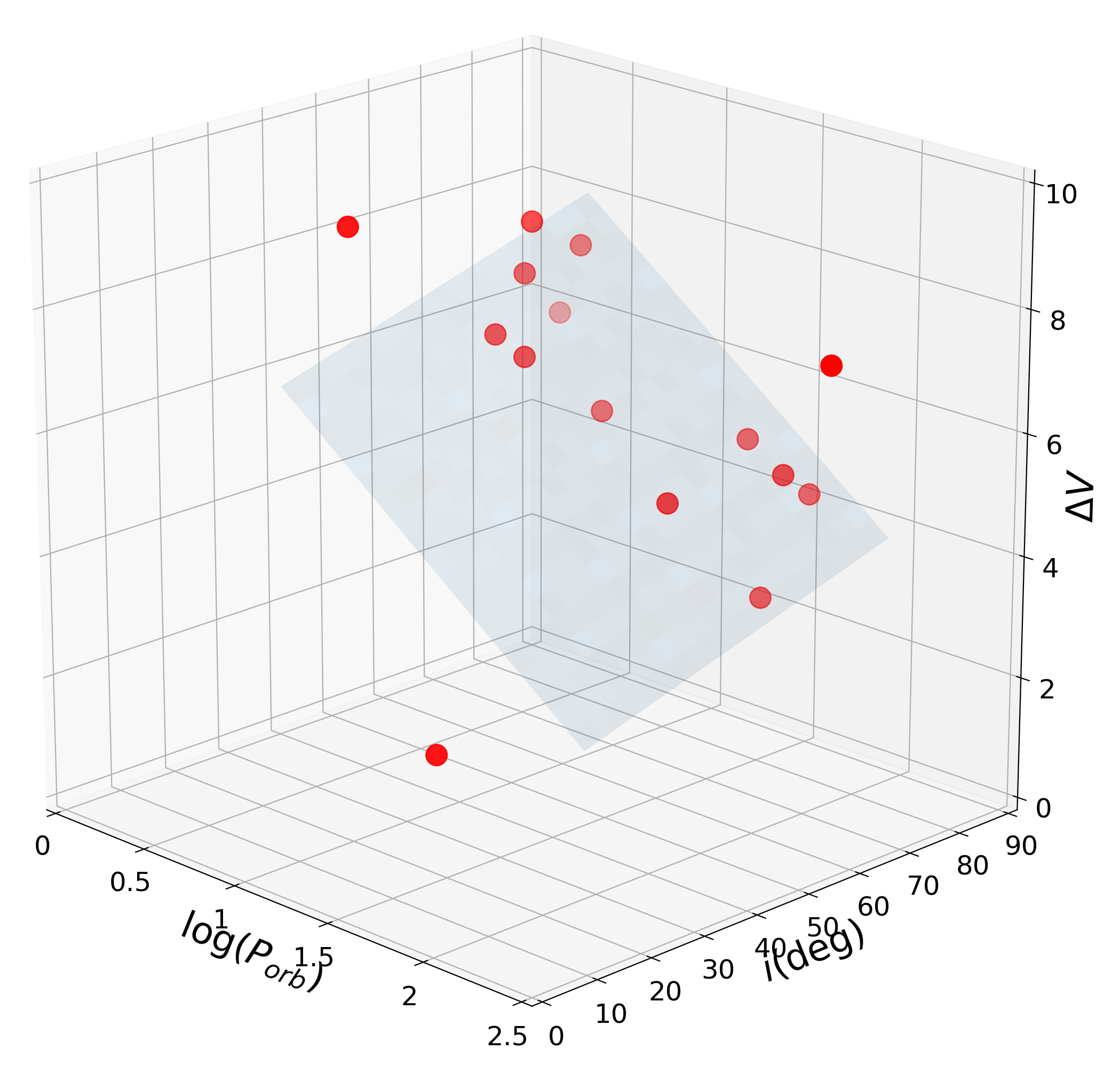}%
        \subcaption{}
        \label{fig:planev}
         \vspace{0.5cm}
    \end{minipage}\hspace{5mm}%
    \begin{minipage}{0.46\textwidth}
        \includegraphics[width=\textwidth]{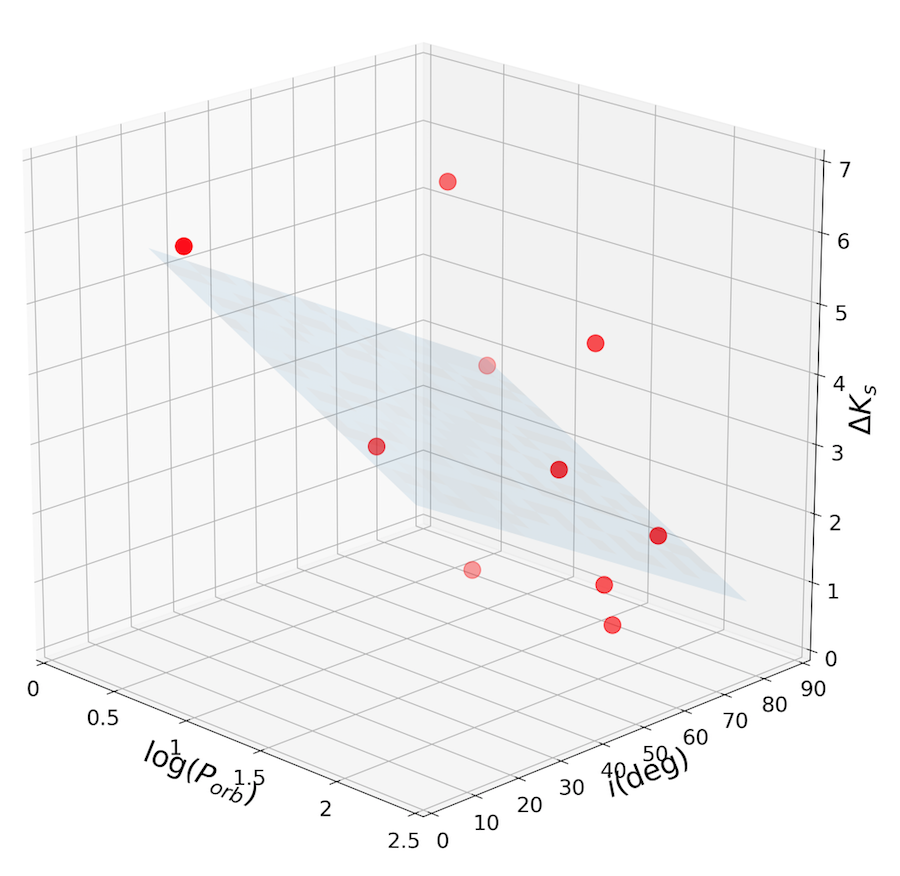}%
       \subcaption{} 
       \label{fig:planek}
        \vspace{0.5cm}       
    \end{minipage}
        \begin{minipage}{0.46\textwidth}
        \includegraphics[width=\textwidth]{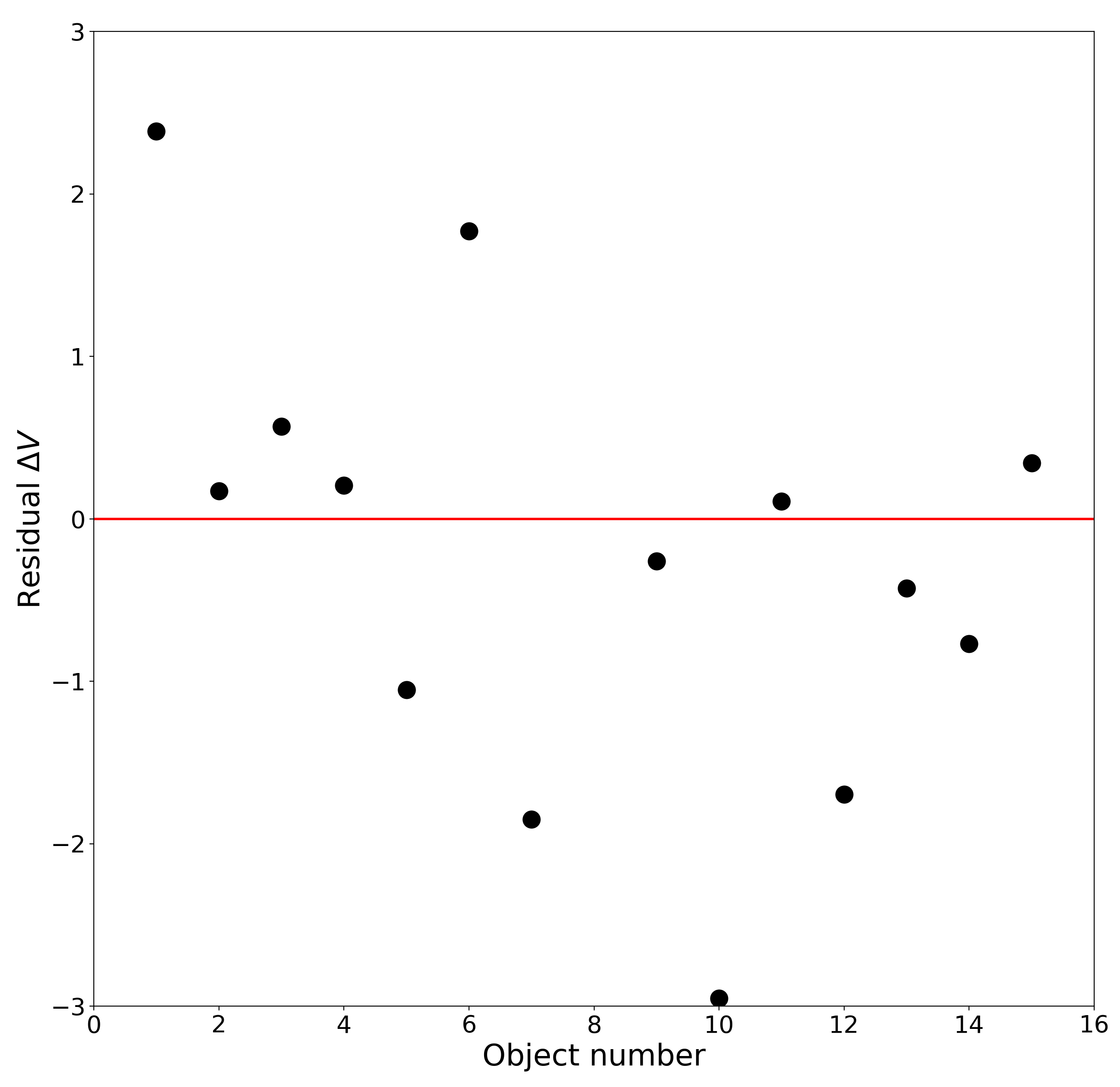}%
        \subcaption{}
        \label{fig:resv}
         \vspace{0.5cm}
    \end{minipage}\hspace{5mm}%
    \begin{minipage}{0.46\textwidth}
        \includegraphics[width=\textwidth]{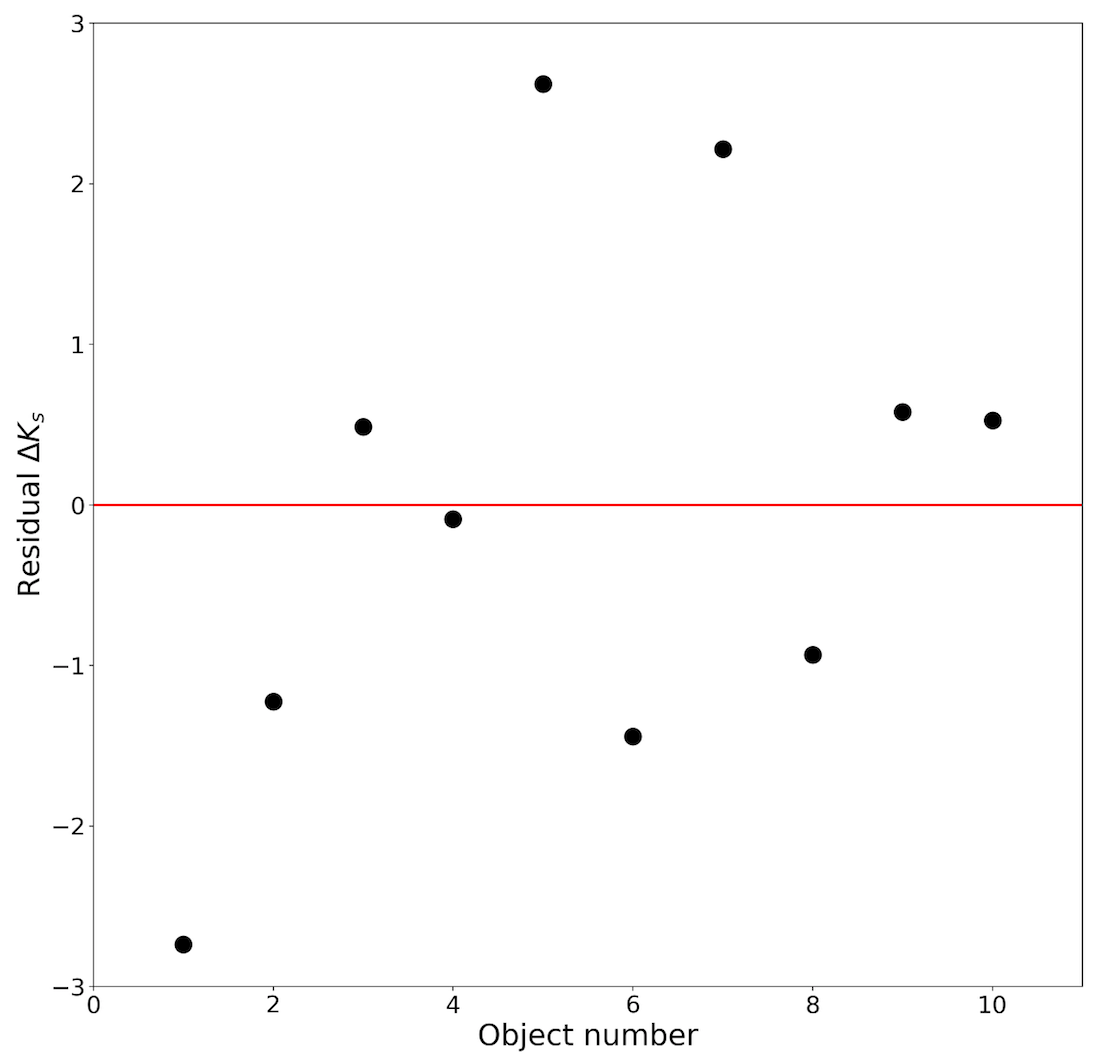}%
       \subcaption{} 
       \label{fig:resk}
        \vspace{0.5cm}       
    \end{minipage}          
	\caption{(a) 3D least squares plane fit to the $V$--band outburst amplitude versus the logarithm of the orbital period $P_{orb}$ and the inclination angle $i$ (for references see Table~\ref{tab:lit}); adapted from \citet{1998MNRAS.295L...1S}. (b) Same as for panel (a) but now for the $K_s$--band amplitude. Residuals of the fit for (c) $\Delta V$ and (d) $\Delta K_s$.}
    \label{fig:planes} 
\end{figure*} 

\begin{table*}
\vspace{5mm}
\begin{center}
\caption{BHXBs with values of outburst and quiescence magnitudes, orbital periods and inclination angles, represented graphically in Figures~\ref{fig:3dV} and~\ref{fig:3dK}. The well known BHXB GRS 1915+105 is not included in this table as it has never been observed in quiescence.}
\label{tab:lit}
\resizebox{\textwidth}{!}{\setlength{\tabcolsep}{15pt}
\begin{tabular}{|lccccc|}
\toprule
 & & \multicolumn{2}{c}{AB Magnitudes} & Orbital & Inclination \\
BHXB & Filter & Outburst & Quiescence & period & angle\\
& &(mag) & (mag) & (h) &($^{\circ}$) \\ 
\midrule
Swift J1357.2-0933 & $K_s$ & 17.4 $\pm$ 0.1$^a$ & 20.09 $\pm$ 0.05$^k$ & 2.8 $\pm$ 0.3$^q$ &$\sim$90$^q$ \\
XTE J1118+480 &$K_s$ & 12.64 $\pm$ 0.09$^b$ & 18.45$^l$ & 4.078414 $\pm$ 0.000005$^r$ &68 $\leq i \leq$ 79$^z$ \\
SAX J1819.3-2525 &$K_s$ & 14.4 $\pm$ 0.3$^c$ & 14.68 $\pm$ 0.05$^m$ & 67.6152 $\pm$ 0.0002$^s$ &72.3 $\pm$ 4.1$^{aa}$\\
XTE J1550-564 &$K_s$ & 13.91 $\pm$ 0.09$^d$ & 18.00 $\pm$ 0.04$^n$ & 37.0088 $\pm$ 0.0001$^n$ &75 $\pm$ 4$^n$\\
GRO J1655-40 &$K_s$ & 14.25$^e$ & 15.12$^o$ & 62.920 $\pm$ 0.003$^t$ &69 $\pm$ 2$^{ab}$\\
GS 2023+338 &$K_s$ & 12.45 $\pm$ 0.03$^f$ & 14.39 $\pm$ 0.05$^p$ & 155.311 $\pm$ 0.002$^u$ &67$^{+3ac}_{-1}$ \\
GX 339-4 &$K_s$ & 12.41 $\pm$ 0.01$^g$ & 15.2 $\pm$ 0.3$^g$ & 42.2088 $\pm$ 0.0001$^v$ & 37 $\leq i \leq$ 78$^v$\\
IGR J17451-3022$\ddagger$ &$K_s$ & 17.39 $\pm$ 0.16$^h$ & 17.59 $\pm$ 0.14$^h$ & 6.284 $\pm$ 0.001$^w$ & 71 $\leq i \leq$ 76$^{ad}$\\
MAXI J1836-194$\ddagger$ &$K_s$ & 14.00 $\pm$ 0.07$^i$ & 20.9 $\pm$ 0.3$^h$ & $<$ 4.9$^x$ & 4 $\leq i \leq$ 15$^x$\\
XTE J1650-500$\ddagger$ &$K_s$ & 15.14 $\pm$ 0.13$^j$ & $>$ 17.8$^h$ & 7.60 $\pm$ 0.02$^y$ & $>$ 47$^y$\\
\midrule
GRO J0422+32* &$V$ & 13.2 $\pm$ 0.1 & 22.4 $\pm$ 0.3 & 5.09 $\pm$ 0.01 & 10 $\leq i \leq$ 50$^{ae}$\\
A 0620-00* &$V$ & 11.22 & 18.32 & 7.7235 $\pm$ 0.0001 & 54.1 $\pm$ 1.1$^{af}$\\
GRS 1009-45* &$V$ & 13.8 $\pm$ 0.3 & 21.67 $\pm$ 0.25 & 6.86 $\pm$ 0.12 & 37 $\leq i \leq$ 80$^{ag}$\\
GRS 1124-68* &$V$ & 13.52 & 20.35 & 10.3825392 $\pm$ 0.0000744 & 39 $\leq i \leq$ 65$^{ah}$\\
H 1705-250* &$V$ & 15.8 $\pm$ 0.5 & 21.5 $\pm$ 0.1 & 12.51 $\pm$ 0.03 & 48 $\leq i \leq$ 80$^{ai}$\\
GS 2000+25* &$V$ & 16.42 & 25.22 & 8.26 &54 $\leq i \leq$ 60$^{aj}$\\
GRO J1655-40* &$V$ & 14.02 & 17.22 & 62.920 $\pm$ 0.003$^t$ & 69$\pm 2^{ak}$\\
GS 2023+338* &$V$ & 11$^{al}$ & 18.44 $\pm$ 0.02 & 155.311 $\pm$ 0.002$^u$ & 67$^{+3ac}_{-1}$\\
GX 339-4 &$V$ & 14.76 $\pm$ 0.02$^{am}$ & 19.67 $\pm$ 0.06$^{an}$ & 42.2088 $\pm$ 0.0001$^v$ & 37 $\leq i \leq$ 78$^v$\\
4U 1543-475 &$V$ & 14.9$^{ao}$ & 16.6 $\pm$ 0.1$^{ap}$ & 26.79377 $\pm$ 0.00007$^{aq}$ & 20.7 $\pm$ 1.5$^{aq}$\\
SAX J1819.3-2525 &$V$ & 8.8$^{ar}$ & 13.96 $\pm$ 0.02$^{as}$ & 67.6152 $\pm$ 0.0002$^{s}$ & 72.3 $\pm$ 4.1$^{aa}$\\
XTE J1118+480 &$V$ & 13.0 $\pm$ 0.3$^{at}$ & 19.6$^{l}$ & 4.078414 $\pm$ 0.000006$^{r}$ &  68 $\leq i \leq$ 79$^{z}$\\
XTE J1550-564 &$V$ & 16.6$^{au}$ & 22.0 $\pm$ 0.4$^{au}$ & 37.0088 $\pm$ 0.0001$^{n}$ & 75 $\pm$ 4$^{n}$\\
GS 1354-64 &$V$ & 16.9$^{av}$ & 21.5$^{aw}$ & 61.068 $\pm$ 0.002$^{aw}$ & $\leq$ 79$^{aw}$\\
XTE J1859+226 &$V$ & 15.3 $\pm$ 0.1$^{ax}$ & 23.3 $\pm$ 0.1$^{ax}$ & 6.58$^{ay}$ & $\leq$ 70$^{ay}$\\
\bottomrule
\multicolumn{6}{l}{{\bf Notes:} Ref: $^a$\citet{2011ATel.3140....1R}, $^b$\citet{2001AstL...27...25T}, $^c$\citet{2003MNRAS.343..169C}, $^d$\citet{2013AA...557A..45C}, $^e$\citet{2005ATel..418....1B},}\\
\multicolumn{6}{l}{$^f$\citet{2015ATel.7663....1G}, $^g$\citet{2002MNRAS.331.1065C}, $^h$this work, $^i$\citet{2011ATel.3619....1R}, $^j$\citet{2012AA...547A..41C}, $^k$\citet{2013MNRAS.434.2696S},}\\
\multicolumn{6}{l}{$^l$\citet{2006ApJ...642..438G}, $^m$\citet{2014ApJ...784....2M}, $^n$\citet{2011ApJ...730...75O}, $^o$\citet{2001ApJ...554.1290G}, $^p$\citet{2004MNRAS.352..877Z},}\\
\multicolumn{6}{l}{$^q$\citet{2013Sci...339.1048C}, $^r$\citet{2004ApJ...612.1026T}, $^s$\citet{2011MNRAS.413L..15C}, $^t$\citet{1998AA...329..538V}, $^u$\citet{1992Natur.355..614C},}\\
\multicolumn{6}{l}{$^v$\citet{2017ApJ...846..132H}, $^w$\citet{2015ATel.7361....1J}, $^x$\citet{2014MNRAS.439.1381R}, $^y$\citet{2004ApJ...616..376O}, $^z$\citet{2013AJ....145...21K},}\\
\multicolumn{6}{l}{$^{aa}$\citet{2014ApJ...784....2M}, $^{ab}$\citet{2002MNRAS.331..351B}, $^{ac}$\citet{2010ApJ...716.1105K}, $^{ad}$\cite{2016A&A...595A..52Z}, $^{ae}$\citet{2000MNRAS.317..528W},}\\
\multicolumn{6}{l}{$^{af}$\citet{2017MNRAS.472.1907V}, $^{ag}$\citet{1996MNRAS.282L..47S}, $^{ah}$\citet{2001AJ....122..971G}, $^{ai}$\citet{1996ApJ...459..226R}, $^{aj}$\citet{2004AJ....127..481I},}\\
\multicolumn{6}{l}{$^{ak}$\citet{2002MNRAS.331..351B}, $^{al}$\citet{2016Natur.529...54K},$^{am}$\citet{2012AJ....143..130B}, $^{an}$\citet{2018MNRAS.478.4710K}, $^{ao}$\citet{1983IAUC.3858....1B},}\\
\multicolumn{6}{l}{$^{ap}$\citet{1998ApJ...499..375O}, $^{aq}$\citet{2003IAUS..212..365O}, $^{ar}$\citet{2001ApJ...555..489O}, $^{as}$\citet{2014ApJ...784....2M}, $^{at}$\citet{2002ApJ...569..423T}, $^{au}$\citet{2002ApJ...568..845O}},\\
\multicolumn{6}{l}{$^{av}$\citet{1990ApJ...361..590K}, $^{aw}$\citet{2009ApJS..181..238C}, $^{ax}$\citet{2002MNRAS.334..999Z} and $^{ay}$\citet{2013RMxAC..42....3C}. $\ddagger$Sources in our sample.}\\
\multicolumn{6}{l}{*Magnitudes and orbital periods are taken from \citet{1998MNRAS.295L...1S} and references therein, unless indicated otherwise.}\\
\end{tabular}}
\end{center}
\end{table*}

\bsp	
\label{lastpage}
\end{document}